\documentclass[useAMS,usenatbib,usegraphicx]{mn2e}
\usepackage{graphicx}                                            
\usepackage{times}
\usepackage{float}
\usepackage{rotating}
\usepackage{epstopdf}
\usepackage{multirow}
\usepackage[para,online,flushleft]{threeparttable}

\def\ltsima{$\; \buildrel < \over \sim \;$}
\def\simlt{\lower.5ex\hbox{\ltsima}}
\def\gtsima{$\; \buildrel > \over \sim \;$}
\def\simgt{\lower.5ex\hbox{\gtsima}}
\def\gsimeq
{\hbox{\raise0.5ex\hbox{$>\lower1.06ex\hbox{$\kern-1.07em{\sim}$}$}}}
\def\lsimeq
{\hbox{\raise0.5ex\hbox{$<\lower1.06ex\hbox{$\kern-1.07em{\sim}$}$}}}

\def\xmm{{\it XMM-Newton }}

\def\xmm{{\it XMM-Newton}}
\def\chandra{{\it Chandra}}

\def\swift{{\it Swift}}

\def\naco{{\it NaCo}}
\def\inte{{\it INTEGRAL}}

\def\apj{ApJ}
\def\mnras{MNRAS}
\def\aap{A\&A}
\def\apjl{ApJ}
\def\apjs{ApJS}
\def\aj{AJ}
\def\ssr{SSR}
\def\araa{ARA\&A}

\def\pasp{PASP}
\def\nat{Nature}
\def\gca{GeCo}

\def\ssv{Swift~J174540.7-290015}
\def\sgr{SGR~J1745-2900}
\def\sgras{Sgr~A$^{\star}$}
\def\axj{AX~J1745.6-2901}

\def\xis{XIS}
\def\xis1{XIS1}
\def\xis2{XIS2}
\def\xis3{XIS3}

\title[] 
 {{\ssv: a new accreting binary in the Galactic Center}}

 \author[G.\ Ponti et al. ]
 {G.~Ponti$^{1}$, 
C.~Jin$^{1}$, 
B.~De Marco$^{1}$, 
N.~Rea$^{2,3}$, 
A.~Rau$^{1}$, 
F.~Haberl$^{1}$, 
F.~Coti Zelati$^{2,4,5}$, 
 \newauthor
E.~Bozzo$^{6}$, 
C.~Ferrigno$^{6}$,
G. C.\ Bower$^{7}$ and 
P. Demorest$^{8}$ 
\\
   $^1$ Max-Planck-Institut f{\"u}r extraterrestrische Physik,  Giessenbachstrasse, 85748, Garching, Germany\\
   $^2$ Anton Pannekoek Institute for Astronomy, University of Amsterdam, Postbus 94249, NL-1090 GE Amsterdam, The Netherlands \\ 
   $^3$ Institute of Space Sciences (CSIC-IEEC), Campus UAB, Carrer Can Magrans s/n, E-08193 Barcelona, Spain \\
   $^4$ Universita dellÕInsubria, via Valleggio 11, I-22100 Como, Italy \\
   $^5$ INAF Ð Osservatorio Astronomico di Brera, Via Bianchi 46, I-23807 Merate (LC), Italy \\
   $^6$ Department of Astronomy, University of Geneva, Chemin dÕEcogia 16, CH-1290 Versoix, Switzerland \\ 
   {$^7$ Academica Sinica Institute of Astronomy and Astrophysics, 645 N. A'ohoku Place, Hilo, HI 96720, USA } \\
   {$^8$ National Radio Astronomy Observatory, P.O. Box O, Socorro, NM 87801, USA } 
    }
\pagerange{\pageref{firstpage}--\pageref{lastpage}}
\usepackage{times}
\begin{document}
\label{firstpage}
 \maketitle
\begin{abstract}
We report on the identification of the new Galactic Center (GC) transient \ssv\ as a likely low mass X-ray binary (LMXB) located at only 16~arcsec from \sgras. This transient was detected on 2016 February $6^{th}$ during the \swift\ GC monitoring, and it showed long-term spectral variations compatible with a hard to soft state transition. We observed the field with \xmm\ on February $26^{th}$ for 35 ks, detecting the source in the soft state, characterised by a low level of variability and a soft X-ray thermal spectrum with a high energy tail (detected by \inte\ up to $\sim50$ keV), typical of either accreting neutron stars or black holes. We observed: i) a high column density of neutral absorbing material, suggesting that \ssv\ is located near or beyond the GC and; ii) a sub-Solar Iron abundance, therefore we argue that Iron is depleted into dust grains. The lack of detection of Fe~K absorption lines, eclipses or dipping suggests that the accretion disc is observed at a low inclination angle. Radio (VLA) observations did not detect any radio counterpart to \ssv. No evidence for X-ray or radio periodicity is found. The location of the transient was observed also in the near-IR with GROND at MPG/ESO La Silla 2.2m telescope and VLT/\naco\ pre- and post-outburst. Within the \chandra\ error region we find multiple objects that display no significant variations.

\end{abstract}


\section{Introduction}

\sgras (Genzel et al. 2010), the supermassive black hole (BH) at the 
Galactic Center (GC), generates a deep gravitational 
potential that is expected to bind tens of thousands of stellar 
remnants, such as neutron stars (NS) or BH that have settled dynamically 
into the central parsec of the Milky Way (Morris 1993; Lee 1995; 
Miranda-Escud{\'e} \& Gould 2000; Muno et al. 2005). 
Deep radio surveys were so far unsuccessful in detecting either ordinary 
or millisecond pulsars, leading to the "missing pulsar problem" (Wharton et 
al. 2012; Dexter \& O'Leary 2014). 
Surprisingly a few years ago the magnetar 
SGR\,J1745-2900 was discovered via its outburst activity (Mori et al. 2013; 
Rea et al. 2013; Eatough et al. 2013; Kaspi et al. 2014; Coti Zelati et al. 2015), 
representing the first radio and X-ray pulsar discovered in the Galatic Center region 
(at less than a few parsecs from \sgras ). Many millisecond radio pulsars 
are expected in the region, but currently not a single one has been detected. 
Due to the three body interaction between this  concentration of degenerate 
stars and multiple stellar systems, a large population of X-ray binaries is 
also expected. The monitoring of X-ray transients allows us to study 
this population (Pfahl \& Loeb 2004; Muno et al. 2005; Hopman 2009; 
Faucher-Giguere \& Loeb 2011).

The majority of accreting X-ray binaries spend most 
of their time in a quiescent state (below the detection limit), sporadically
interrupted by outbursts during which the X-ray luminosity rises 
by many orders of magnitudes (Fender \& Belloni 2012). 
Pioneering works on monitoring of GC X-ray sources have been performed 
with \chandra\ and \swift\ (Muno et al. 2005; Wijnands et al. 2006; 
Campana et al. 2009; Degenaar et al. 2012; 2015). 

\ssv\ was discovered by the \swift\ satellite on Feb. 6$^{th}$ 2016 (Reynolds 
et al. 2016) during the first X-ray observation after the GC exited the Solar 
constraint window. The \swift\ X-ray telescope (XRT) data showed a bright 
($F\sim2\times10^{-10}$~erg~cm$^{-2}$~s$^{-1}$, in the $2-10$ keV range) transient 
located at less than 16~arcsec from \sgras\ (Reynolds et al. 2016). 
\ssv\ was detected also in the 20-80~keV energy range (Esposito et al. 2016), 
while observations with the Karl G. Jansky Very Large Array (VLA) 
and Giant Metrewave Radio Telescope (GMRT) did not detect the radio 
counterpart
(however the limits still leave open the possibility of a hard state accreting 
BH or NS) or any pulsation (Maan et al. 2016; Bower et al. 2016).
Within a week from this discovery, \chandra\ observed the field and refined 
the transient position to: RA$_{\rm J2000}=$17:45:40.66$\pm$0.34, 
Dec$_{\rm J2000}=$-29:00:15.61$\pm$0.33, where no previous X-ray sources 
were ever detected (Baganoff et al. 2016). Therefore, \ssv\ is a new transient 
source. 

In this paper we report the identification of the source as a low-inclination 
X-ray binary hosting a neutron star or a black hole, most probably with 
a low mass companion. We discuss the temporal and spectral characteristics 
of the transient using the long-term daily X-ray monitoring of the field 
performed by the \swift\ satellite, the high energy long term 
monitoring provided by the \inte\ satellite, a 35~ks \xmm\ 
observation obtained in Director's Discretionary Time (to study the dust 
scattering halo) performed about 20 days after the onset of the outburst 
as well as VLA observations. Finally, we report on the search 
for the near-IR counterpart within VLT/\naco\ and GROND data. 

\section{X-ray observations and data reduction}

\subsection{XMM-Newton} 
\xmm\ (Jansen et al. 2001) observed \ssv\ on February 26$^{th}$ for 35~ks.
We processed this data set starting from the observation data files. 
The data have been treated using the latest 
version (15.0.0) of the \xmm\ Science Analysis System ({\sc sas}) with 
the most recent (13 March 2016) calibrations. We used all EPIC (Str\"uder et 
al. 2001; Turner et al. 2001) cameras equipped with the medium filter.
The arrival times of the events were corrected to the Solar system 
barycentre, applying the {\sc barycen} task of {\sc sas}. 
We screened the periods of enhanced particle-induced activity by inspecting 
the EPIC-pn 10-15 keV light curves, binned with 5~s resolution, extracted from 
a circular region, with 5~arcmin radius, close to the border of the field of view. 
We then filter out all the periods with more than 0.85 cts~s$^{-1}$. 

Since the main focus of the observation was the study of the full extension of 
the dust scattering halo around the source, the EPIC-pn camera 
was in Full Frame mode. Having a frame time of 73.4~ms, 
the inner core of the point spread function shows pile up. On the other hand, 
both EPIC-MOS1 and MOS2 were in timing mode, with a frame time 
of 1.75~ms. This assures that no pile up is present in these datasets. 
A bad column is falling in the center of the extraction region in the 
EPIC-MOS1 camera, inducing lower statistics. Therefore these data 
have been used for consistency checks only. 
To reduce pile up, the source photons in the EPIC-pn camera were 
extracted from an annular region. The outer radius of the annulus 
was limited to 40 arcsec, to avoid contribution from the low mass 
X-ray binary \axj\ (Ponti et al. 2015) and the inner 
radius has been fixed to 15 arcsec. By comparing this spectrum from 
the wings of the Point Spread Function with the pile-up free EPIC-MOS2 
spectrum, we verified that it is not affected by pile up. 
We note that the magnetar SGR~J1745-2900 (Rea et al. 2013) is located 
inside the source extraction region. However, the magnetar is currently 
observed at a 2-10~keV flux of $F_{\rm 2-10}=3.8\times10^{-13}$ 
erg s$^{-1}$ cm$^{-2}$, about three orders of magnitude lower than 
the instantaneous 2-10~keV flux of \ssv\ ($F_{\rm 2-10}=2.9\times10^{-10}$
erg s$^{-1}$ cm$^{-2}$). Therefore, SGR~J1745-2900 gives a negligible 
contribution to the light curves and spectra of \ssv.

A strong contribution from the emission of the supernova remnant Sgr~A~East 
(Maeda et al. 2002; Ponti et al. 2015) is expected within the EPIC-pn  
extraction region of \ssv, therefore we extracted the background plus diffuse 
emission from a long archival observation ({\sc obsID} 0505670101) 
during which \axj\ had flux and spectral properties similar to the ones during 
this observation (being in the soft state and at similar flux), the magnetar 
\sgr\ was in quiescence and \sgras\ did not show any flare (Ponti et al. 2015). 
In the EPIC-MOS2 camera, the source photons were extracted between 
{\sc rawx} $285$ and $330$. The larger extraction region, imposed by 
the timing mode, implies a larger contribution from \axj\ and diffuse 
emission, compared to the EPIC-pn spectrum. We converted the source 
extraction region (from {\sc rawx} coordinates into sky coordinates 
through the {\sc sas} task {\sc ecoordconv}) and we extracted the background 
plus diffuse emission from the EPIC-MOS2 camera, during the same long 
observation ({\sc obsID} 0505670101) with \ssv\ in quiescence. 
Because of the higher contamination by \axj\ we use the EPIC-MOS 
spectra for comparison only. We note that the same best fit parameters
(with all values consistent with the ones obtained by fitting the EPIC-pn 
spectrum; \S \ref{specXMM}) were obtained by fitting the EPIC-MOS spectra. 

The combination of high column density of absorbing material 
($N_H=18.7\times10^{22}$~cm$^{-2}$) and the intense diffuse emission
in the direction towards the Galactic center, prevents us from obtaining 
a good characterisation of the source emission below $\sim2$~keV 
(e.g., the low energy absorption edges).
For this reason in this work we do not consider the data from the Reflection 
Grating Spectrometer (RGS) instruments.

Spectral fitting was performed using {\sc XSpec} version 12.8.2. The spectra 
were grouped so that each bin contains at lest 30 counts. 

\subsection{Swift} 

The {\it Swift} X-ray Telescope (XRT) has been monitoring the central 
$\sim15$ arcmin of the Galaxy daily since year 2012 (Degenaar et al. 2015). 
We downloaded all the {\it Swift} PC mode datasets between 5-2-2016 
and 10-3-2016 from the {\sc heasarc} data archive. This comprises 31 
PC mode XRT observations. 
To characterise the contribution from the background plus complex diffuse 
emission within the extraction region of \ssv\ (that contains a contribution 
from SGR~J1745-2900 and \axj), we used the \swift\ data before the 
outburst of \ssv. Since there was no {\it Swift} observation 
of this region between 03-11-2015 and 05-02-2016, we retrieved 15 {\it Swift} 
observations between 15-10-2015 and 02-11-2015 in XRT PC mode when 
the source was quiescent. These 15 observations are used to measure 
the background plus diffuse emission underneath \ssv.

We applied the standard data reduction following the threads described 
on the {\it Swift} website. Because of the high absorption column density 
towards the source, we considered only data in the 2-10~keV band.
Images and spectra were produced using the 
{\sc xselect } (v2.4c) package. Since \ssv\ is very bright,
significant pile up is affecting all the PC mode observations. We followed 
the standard XRT pile-up thread to determine the pile-up level. 
We extracted the products from an annulus with inner radius 
of 15$\arcsec$, to ensure that spectra, fluxes and light curves are not affected 
by pile up. A 40$\arcsec$ outer radius is used to include most source flux and 
less contamination from \axj. 
We manually adjusted the astrometry of all datasets by visually inspecting 
the images of all the observations. This has been obtained by 
adjusting the relative position between \ssv\ and AX J1745.6-2901. 
This allows us to correct the astrometry shifts, which would otherwise
affect the PSF correction and the correct determination of the source flux. 
By comparing the light curve with that derived without PSF correction,
we ensured that the uncertainties in the PSF correction are not the cause 
of the observed source variability. We also note that the background 
flux underneath \ssv\ is only 3\% of the source flux in the 2-10 keV band, 
so it does not significantly contaminate the light curve of \ssv, either.

In addition to the PC mode data, there are also 12 {\it Swift} XRT 
Windowed Timing (WT) mode observations of \ssv. 
Compared to the PC mode, the WT mode provides higher time resolution 
and more counts, but the 1-dimensional image in the WT mode makes 
it difficult to determine the background underneath the source region 
from previous observations as we did in the PC mode. However, since 
the background flux is only 3\% of the source flux, we use anyway the WT 
mode observations as an independent check of the PC mode results.

\begin{figure}
\centering
\includegraphics[scale=0.35]{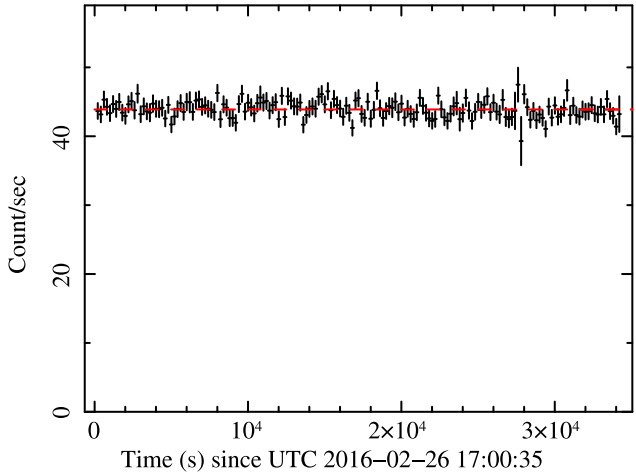}
\caption{PSF corrected EPIC-pn light curve of \ssv\ in the 2-10 keV band. 
The light curve is extracted from an annular region with inner and outer 
radii of $15^{\prime\prime}$ and $40^{\prime\prime}$, respectively,  
and binned at 200~s. No bursts (not even in 1~s binned 
light light curves) or dips are observed. The red dashed line shows 
the best fit with a constant (that results in 43.9 cts s$^{-1}$ and 
$\chi^2=161.7$ for 170 dof). 
}
\label{cluce}
\end{figure} 
\ssv\ is not piled up in the WT mode, so we defined a circular source region 
of 20$\arcsec$ radius, centred on the pixel which is the perpendicular 
projection of the source position on the 1-dimensional WT mode image 
(again we consider only data in the 2-10~keV band). 
The choice of 20$\arcsec$ radius was a compromise between including 
more source flux and less contamination from \axj. 
For eleven WT mode observations, there were also PC mode observations 
on the same day, for which we could compare the spectra between the 
two modes. For nine observations, both the spectral shape and normalization 
were consistent between the two modes, except that WT mode provided 
more counts and thus higher S/N. The remaining two WT mode spectra 
(ObsID: 00035063134, 00035063136) have the same spectral shape as
in the PC mode but with different normalisation. We found that this is caused 
by the bad pixels within the source extraction region in these two WT mode
observations, which affected the source flux and PSF correction, but did not 
affect the spectral shape. We excluded these two observations from further 
analysis. Since the source spectra from WT mode observations were consistent 
with the PC mode, and the background underneath \ssv\ was much weaker 
than the source itself, we decided to apply the same background flux and 
spectra for the WT mode data as for the PC mode, and this should not affect 
our results.

The exposure maps and ancillary files for the spectra were computed with the
{\sc xrtexpomap} and {\sc xrtmkarf} tools provided in the {\sc heasoft} (6.18) 
software. 

\subsection{\inte}

The outburst of the source was observed by \inte\ starting from 2016 
February 11, i.e. during satellite revolution 1643. At the time of writing, only 
consolidated \inte\ data up to revolution 1652 were made available and thus 
we were able to observe the evolution of the source hard X-ray emission 
until 2016 March 6. 
As the source is located right into the Galactic center, which is a complex and 
crowded region, we also made use of the last data collected toward this region 
before the outburst of \ssv. These included data during revolutions 1597, 
1603, and 1604, covering the period from 2015 October 11 to 19. 

\inte\ observations are divided into ``science windows'' (SCWs), i.e.,  pointings 
with typical durations of $\sim$2-3~ks. We considered only SCWs where 
the source was within a maximum off-axis angle of 12~deg from the satellite 
aim point, in order to reduce the uncertainties on the energy calibrations. 
All data were analyzed by using version 10.2 of the Off-line Scientific Analysis 
software (OSA) distributed by the ISDC (Courvoisier et al. 2003). 
We first extracted the IBIS/ISGRI (Ubertini et al. 2003; Lebrun et al. 2003) 
mosaics in the 20-100~keV energy band and the JEM-X (Lund et al. 2003) 
mosaics in the 3-10~keV energy band by summing up all data in revolutions 
1597 to 1604. No source was detected at a position consistent with \ssv\ 
in the 20-100~keV energy band and we estimated a 3$\sigma$ upper limit 
on the corresponding energy range of 3~mCrab (corresponding\footnote{The 
conversion between the upper limit on the source count-rate and flux has 
been done by using the observations of the Crab in revolution 1597, 
as described in Bozzo et al. (2016).} to roughly 
5.4$\times$10$^{-11}$~erg~cm$^{-2}$~s$^{-1}$). 
From the JEM-X mosaic, we noticed that the well known transient LMXB 
AX\,J1745.6-2901 (Ponti et al. 2015) was detected at a position consistent 
with that of \ssv. The limited spatial resolution of the JEM-X instrument 
does not allow us to disentangle the emission of AX\,J1745.6-2901 
from the one of \ssv. 
Therefore, we decided to extract a JEM-X spectrum of AX\,J1745.6-2901 
from observation taken before the outburst of \ssv. We then used this file 
as a background for the spectrum of \ssv, that we obtained from the 
later observations (we checked in the \swift\ data that \axj\ did not go 
through major flux and/or state changes during the considered period). 
The JEM-X spectrum of AX\,J1745.6-2901 could be well fit 
($\chi^2_{\rm red}$/d.o.f.=0.5/5) with a power-law model of photon index 
$\Gamma=2.2\pm0.5$ 
($F_{3-10~keV}\sim3.5\times10^{-10}$~erg~cm$^{-2}$~s$^{-1}$)\footnote{Note 
that in order to avoid contamination issues both the JEM-X and ISGRI 
spectra used in the present work were extracted from the instrument 
mosaics with the {\sc mosaic\_spec} tool rather than running the OSA 
software down to the SPE level (see discussion in Bozzo et al. 2016).}. 

\begin{figure*}
\hspace{-0.4cm}
  \includegraphics[width=15.1cm]{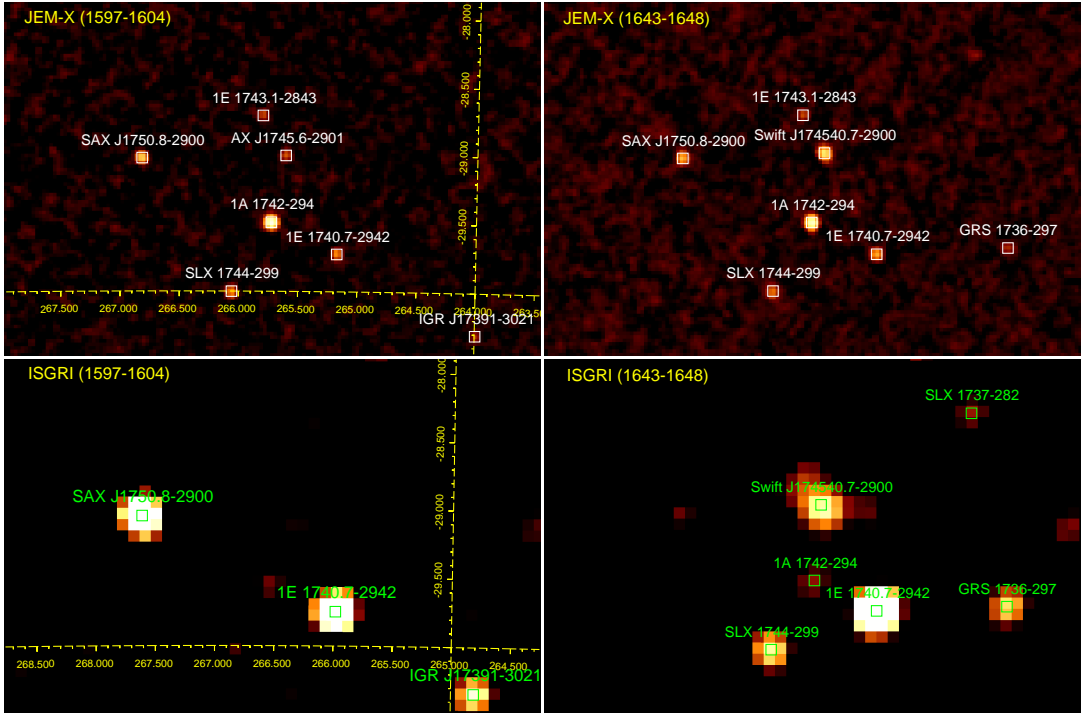}
  \caption{ Mosaiked JEM-X (3-10~keV) and IBIS/ISGRI (20-40~keV) images 
  around the position of \ssv\ obtained from the revolutions 1597-1604 performed 
  before the source outburst (top and bottom left) and the revolutions 
  1643-1648 performed during the first part of the source outburst (top and 
  bottom right). As it can be seen, \ssv\ is only detected from revolution 
  1643 onward in IBIS/ISGRI, while in JEM-X AX\,J1745.6-2901 is detected 
  in the earlier revolutions at a position consistent with \swift\ (albeit at 
  a much lower luminosity (see text for details).} 
  \label{fig:mosaic}
\end{figure*}

For all \inte\ revolutions from 1643 to 1652, we built the IBIS/ISGRI and 
JEM-X mosaics and extracted the source spectra in 13 energy bins 
for ISGRI and 8 energy bins for JEM-X (we used data from both the JEM-X1 
and JEM-X2 units). The JEM-X spectrum obtained during revolution 1643 
revealed a 3-20~keV flux of 
$F_{3-20~keV}=1.0\times$10$^{-9}$~erg~cm$^{-2}$~s$^{-1}$, 
therefore AX\,J1745.6-2901 contaminated the emission from \ssv\ by 
no more than 30\% (see also Fig.~\ref{fig:mosaic}). 
As AX\,J1745.6-2901 was not detected in the previous IBIS/ISGRI mosaic, 
the recorded 
20-100~keV X-ray flux of 7.4$\times$10$^{-10}$~erg~cm$^{-2}$~s$^{-1}$ is to 
be completely attributed to \ssv. This object remained bright in all \inte\ data 
we used, and was detected at high significance in the IBIS/ISGRI and JEM-X 
mosaics of all revolutions. 

\subsection{Karl G. Jansky Very Large Array}

VLA observations of \ssv\ were carried out on two epochs, 25 Feb 2016  and 
25 Mar 2016. As we will discuss later (see \S \ref{secSwift}), \ssv\ was in the soft 
state during at least the first of these VLA observations. 
In each epoch, data were obtained in two frequency bands, 
C and Ku, with center frequencies of 6.0 and 15.0 GHz, respectively.
The VLA correlator was simultaneously configured for imaging and a phased 
array beam centered on \ssv.  The VLA was in its C configuration for
both observations with a maximum baseline length of 3 km.

Imaging data were obtained over frequency ranges of 3.976 -- 8.072 GHz 
and 12.952 -- 17.048 GHz, respectively. The correlator was configured 
to produce cross-correlations that were split into 64 frequency windows each 
with 128 channels in dual circular polarization.
These imaging data were reduced using the CASA VLA pipeline. 
The radio galaxy 3C 286 was used to set the flux calibration and the compact
source J1744-3116 was used as a phase calibration source for \ssv. 
Additional steps of self-calibration with \sgras\ as the reference were performed 
using only visibilities with baselines longer than 50 $k\lambda$ and 20 $k\lambda$ 
in Ku and C bands, respectively.
Images of the \ssv\ field using only these long baseline visibilities were produced. 
These long baselines filtered out much of the substantial emission associated 
with the Northern arm of Sgr A West. Details of the epochs and images are 
summarized in Table~\ref{tab:vla}. No radio source was identified at
the location of \swift\ in either epoch at either frequency.
\begin{table*}
\begin{tabular}{ l l r l r }
\hline
\hline
Epoch & UT & Center Frequency & Beam & RMS \\
& & (GHz) &  & (mJy) \\
\hline
25 Feb 2016 & 16:28 -- 17:09 UT & 6.0 & $7.8\arcsec \times 2.3\arcsec$, 22$^\circ$ & 53 \\
\dots       & \dots            & 15.0 & $4.3\arcsec \times 0.9\arcsec$, 29$^\circ$ & 6  \\
25 Mar 2016 & 13:45 -- 14:45 UT & 6.0 & $6.3\arcsec \times 2.2\arcsec$, 18$^\circ$ & 53 \\
\dots       & \dots            & 15.0 & $3.5\arcsec \times 0.9\arcsec$, 22$^\circ$ & 14  \\
\hline
\end{tabular}
\caption{VLA Image Summary.} 
\label{tab:vla}
\end{table*}

We also recorded high time resolution data from a phased array beam at 
the position of \ssv. In both frequency bands, these data were taken 
with 4.096 GHz total bandwidth divided into 1024 frequency channels 
at 250 $\mu$s time resolution.  We searched for dispersed, periodic signals 
over dispersion measures ranging from 0 to 10,000 pc/cm$^{-3}$ and pulse 
frequency drift (source acceleration) up to $5 \times 10^{-5}$ Hz/s. 
At both frequencies, pulses from the bright, nearby 3.76-s magnetar 
PSR J1745-2900 were detected with high signal to noise ratio, but no 
other significant periodicities were found. 
Assuming 10\% pulse duty cycle, the 10-sigma flux density limits 
for the periodicity searches were 30 $\mu$Jy at C-band and 
45 $\mu$Jy at Ku-band.

\

Throughout the paper we assume a distance to the source equal 
to 8~kpc (Genzel et al. 2010; Gillessen et al. 2013). 
Error-bars are quoted at the 90 per cent confidence level for a single 
parameter of interest. 
We also assume that the accretion disc is observed face on. 
This is in agreement with the indication (\S 3.1) that the system is observed 
at low inclination ($i<60^\circ$; Frank et al. 1987; Diaz-Trigo et al. 2006; 
Ponti et al. 2016). Would the inclination be $i=60^\circ$, the inner 
disc radii and luminosities would be 1.4 and 2 times larger, respectively.  

\section{Short term variability} 
\subsection{The \xmm\ light curve}

Figure \ref{cluce} shows the PSF corrected EPIC-pn light curve 
in the 2-10~keV band with 200~s bins. 
No variability is observed and the data are well fit 
with a constant that is shown by the red dashed line 
($\chi^2=161.7$ for 170 degrees of freedom; dof). 
No bursts or dipping events are observed during the \xmm\ observation
(White \& Mason 1985; Frank et al. 1987; Diaz-Trigo et al. 2006). 
To refine our capability to detect bursts and dips we investigated a variety 
of energy bands and time bins. No bursts are detected in light curves with 
time bins, as short as sub-second. We also investigated soft energy bands, 
where the effect of dipping is stronger (Diaz-Trigo et al. 2006; Ponti et al. 2016), 
and their hardness ratio, but no dipping events were found.

\subsection{Timing analysis: broad band noise}

We extracted MOS and pn light curves in the energy bands 2-10 keV, 
2-6 keV, and 6-10 keV, with a time resolution of 3.5 ms and 200 ms, 
respectively. 
The power spectral density function (PSD) was computed in each energy 
band and for each instrument separately. The Poisson noise level was 
estimated from the mean power at frequencies $>50$ Hz, where counting 
noise variability dominates. The relatively long frame time of the EPIC pn 
instrument in Full Frame observing mode prevents us from sampling 
these frequencies. Therefore, the Poisson noise level cannot be 
accurately estimated, thus we did not consider the EPIC-pn PSD further.

\begin{figure}
\centering
\includegraphics[scale=0.46]{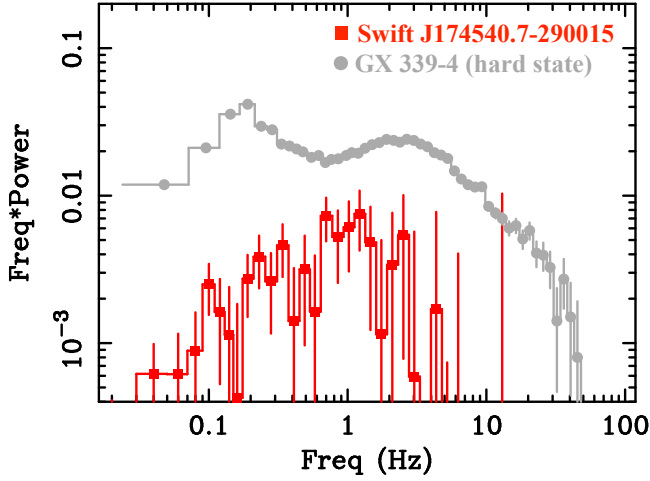}
\caption{The PSD (in units of $[rms/mean]^2$) of Swift J174540.7-290015 
(red squares), computed from MOS2 data in the 2-10 keV energy band. 
The 2-10 keV PSD of GX 339-4 in a high-luminosity hard state 
(De Marco et al. 2015) is reported for comparison (gray dots). 
}
\label{PSD}
\end{figure}

Figure \ref{PSD} shows the PSD of the MOS data in the 2-10 keV band. 
For comparison we also plot the PSD of the BH X-ray binary 
GX 339-4 in a high-luminosity hard state (De Marco et al. 2015). 
SwiftJ174540.7-290015 displays significantly less power than observed 
in a typical hard state of a BHXRB. Indeed, the 2-10 keV fractional 
root-mean-square (rms) variability amplitude (e.g. Nandra et al. 1997; 
Vaughan et al. 2003; Ponti et al. 2004) in the 0.1-64 Hz frequency interval 
is estimated to be $0.07\pm0.02$. 
This value is consistent with what is typically observed in either a BH 
or NS XRB in a soft/soft-intermediate state (Mu\~noz-Darias et al. 2011; 
2014).

\subsection{Timing analysis: search for periodic signals}

We searched for periodic signals using the \xmm\ pn and MOSs
data, as well as \swift\ WT-mode data (being sensible to signals between
the Nyquist frequency, and the frequency resolution of each
dataset; van der Klis 1988). We used a modified version of the {\tt
Xronos} analysis software to search for periodicities following the
prescriptions described in Vaughan et al. (1994), and we did not find
any periodic or quasi-periodic signal in any of the reported X-ray
observations (we had accounted for the number of bins searched, and the
different d.o.f. of the noise power distribution in the non-detection
level). We have searched all dataset performing Fast Fourier 
Transforms over the total length of the observations, but also over 
small intervals of 0.5, 1, 3, and 5\,ks in order to search for signals 
that might have been possibly smeared by Doppler shifts due 
to hour-long orbital periods. Unfortunately, the dataset with by far 
the largest number of counts, the EPIC-pn, had a timing resolution 
of $\sim73$\,ms, making our searches rather insensitive to fast 
spinning pulsars (i.e. with spin periods of $\sim$ twice this value).

\begin{figure}
\centering
\includegraphics[scale=0.3,angle=90]{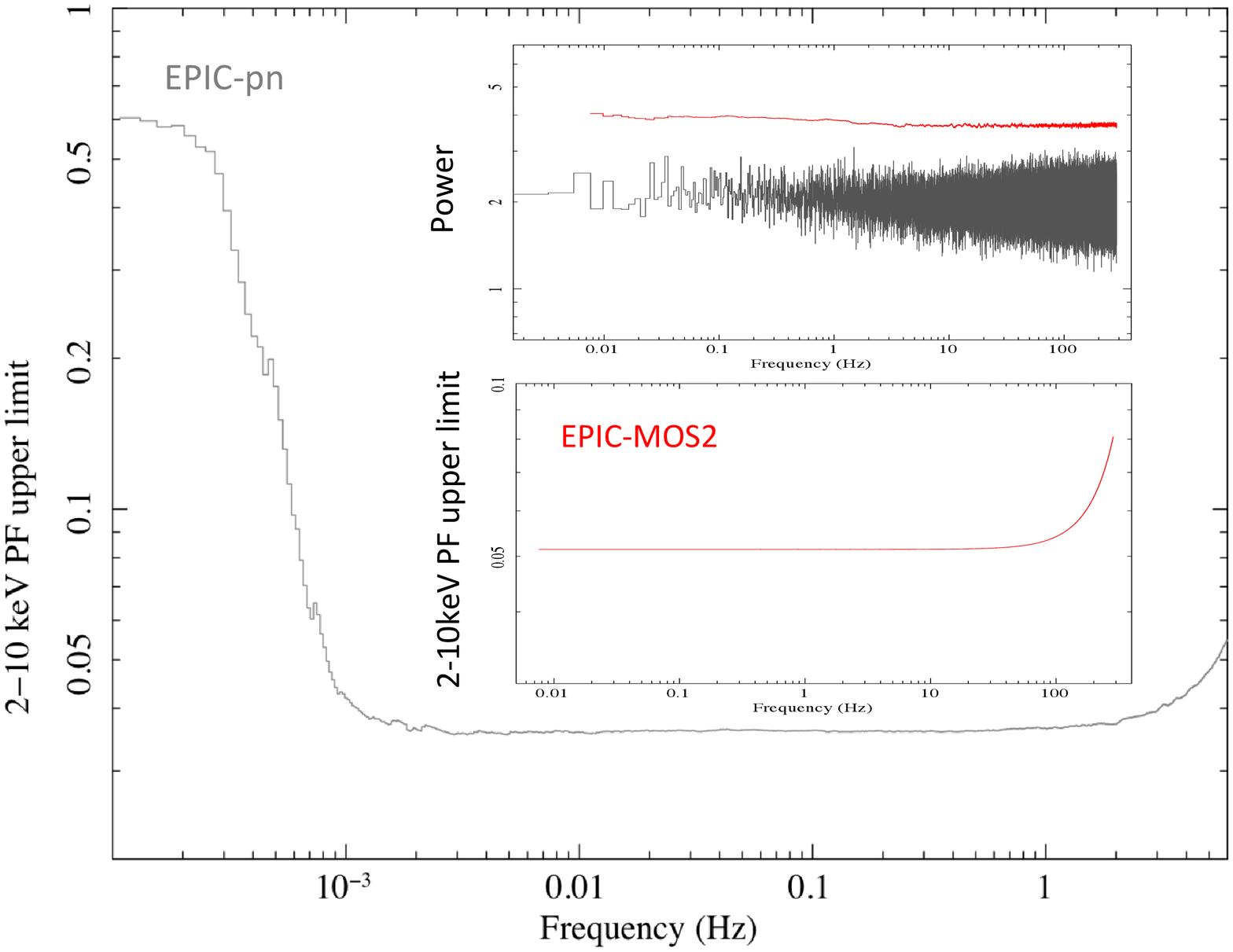}
\caption{{\it (Main panel)} Pulsed fraction 3$\sigma$ upper limits 
as derived for the 2-10keV EPIC-pn. The fractional PF is in units 
of 100~\%. {\it (Upper inset)} MOS2 power 
density spectra. {(\it Lower inset)} Fractional pulsed fraction 3$\sigma$ 
upper limits as derived for the 2-10keV MOS2.}
\label{PF}
\end{figure}

We computed the 3$\sigma$ upper limits on the sinusoid semi-amplitude
pulsed fraction ($PF$) according to Vaughan et al. (1994)
and Israel \& Stella (1996). The deepest limits at frequencies $<
6$\,Hz were derived from the EPIC-pn data, having the larger number of
collected photons. Note that given the
nature of this source, if a very short orbital period causes Doppler 
smearing of the putative signal (i.e. if the signal is not strong enough 
to be detected in our small $\sim$1\,ks chunks that might be free of 
Doppler smearing), these PF limits are not constraining. To have a 
handle on the PF limits on faster periodicities (anyway smaller 
than 300\,Hz), we used the MOSs data (taken in timing mode: 
1.75\,ms timing resolution) as well as the \swift\ WT data (1.7\,ms 
timing resolution), although having to cope with a much reduced 
number of counts, and again using the full length of the observation 
and searching in small consecutive time intervals. We show the 
resulting 3$\sigma$ upper limit on the PF of a periodic signal (if not 
Doppler smeared in the presence of a short orbital period) for both 
pn and MOS2 data in Fig. \ref{PF}.

\section{Spectral analysis of the deep \xmm\ observation} 
\label{specXMM}
\subsection{Characterisation of the mean spectrum}
\label{sec-xmm-meanspec}

We fit the EPIC-pn spectrum by always including an absorption 
component from neutral matter plus the contribution from the dust 
scattering halo. The neutral absorption is fitted with the model 
{\sc TBnew} (see Wilms et al. 2000; 2011) with cross sections of Verner 
et al. (1996) and abundances of Wilms et al. (2000). The dust scattering 
halo is fitted with the model {\sc dust} (Predehl \& Schmitt 1995). We assume 
that the dust scattering optical depth at 1~keV is tied to the X-ray absorbing 
column density by the relation: $\tau=0.324\times(N_H/10^{22} cm^{-2}$; 
as in Nowak et al. 2012). However, we note that a more proper 
modelling of the dust scattering effect on the observed spectrum
requires better knowledges about the location and properties of the dust
along the line of sight (Smith, Valencic \& Corrales 2016), 
as well as a better constrained location of \ssv, which would
then require detailed analysis of the extended emission from
the dust scattering halo. This is already beyond the scope of this paper,
but will be included in our further study.

We first fit the spectrum with a series of single component models. 
We start with an absorbed power law model 
({\sc dust*tbnew*powerlaw} in {\sc xspec}). Large residuals remain at 
high energy ($\chi^2=1530.9$ for 1189 dof; see Tab. \ref{single}). 
Moreover, the best fit power law spectral index is unrealistically steep 
($\Gamma=5.76\pm0.04$), clearly indicating the thermal nature of this emission. 
Therefore, we fit the spectrum with either an absorbed black body 
({\sc dust*tbnew*bbody}) or an absorbed multi-temperature disk black 
body component ({\sc dust*tbnew*diskbb}). 
None of these two models provides an acceptable 
fit, all leaving significant residuals at high energies (see Tab. \ref{single}).
\begin{figure}
\hskip -0.2cm
\includegraphics[scale=0.35,angle=-90]{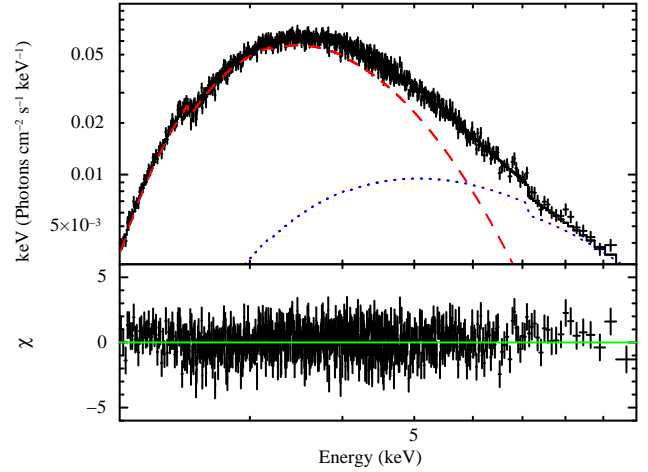}
\caption{{\it (Upper panel) } EPIC-pn mean spectrum fitted with a black 
body plus thermal Comptonization component, all absorbed 
by neutral material. The red dashed and blue dotted lines 
show the contribution from the black body and thermal 
Comptonization components, respectively (see Tab. \ref{double}). 
{\it (Lower panel)} Residuals compared to the best fit model. }
\label{Spec}
\end{figure}

\begin{table} 
\begin{center}
\begin{tabular}{ c c c c c c }
\hline
\hline
\multicolumn{2}{c}{Single component model} \\
\hline
\hline
\multicolumn{2}{c}{\sc dust*tbnew*powerlaw} \\ 
\hline
$N_H$           & $25.6\pm0.2$        \\
$\Gamma$    & $5.76\pm0.04$      \\
$A_{PL}$       & $128\pm8$            \\
$\chi^2/dof$  & 1530.9 / 1189        \\
\hline
\multicolumn{2}{c}{\sc dust*tbnew*bbody} \\ 
\hline
$N_H$          & $13.7\pm0.2$         \\
$kT_{BB}$     & $0.693\pm0.003$   \\
$A_{BB}$   & $0.0152\pm0.0003$   \\
$\chi^2/dof$  & 2897.9 / 1189          \\
\hline
\multicolumn{2}{c}{\sc dust*tbnew*diskbb} \\
\hline
$N_H$          & $15.9\pm0.1$       \\
$kT_{DBB}$  & $0.832\pm0.006$ \\
$A_{DBB}$    & $278\pm14$          \\
$\chi^2/dof$  & 2479.3 / 1189        \\
\hline
\end{tabular}
\caption{Best fit parameters of the EPIC-pn and MOS2 spectra 
with a single component models. Column densities ($N_H$) are 
in units of $10^{22}$~cm$^{-2}$, temperatures ($kT_{BB}$ and 
$kT_{DBB}$) are in units of keV, the power law normalisation ($A_{PL}$) 
is units of photons keV$^{-1}$ cm$^{-2}$ s$^{-1}$ at 1 keV, the disc 
black body normalisation ($A_{DBB}$) is in units of 
$(R_{in}/km)^2/(D/10~kpc)^2$, where $R_{in}$ is the apparent inner disc 
radius (in km units), $D$ is the distance to the source (in 10~kpc units) 
and the black body normalisation ($A_{BB}$) is in units of $L_{39}/(D/10~kpc)^2$, 
where $L_{39}$ is the source luminosity in units of $10^{39}$~erg$^{-1}$. }
\label{single}
\end{center}
\end{table} 
\begin{table*} 
\begin{center}
\begin{tabular}{ c c c c c }
\hline
\hline
\multicolumn{5}{c}{Double component model~{\sc abs = dust*tbnew}}\\
\multicolumn{2}{c}{\sc abs*(bbody+nthComp)} & \multicolumn{1}{c}{\sc abs*(diskbb+nthComp)} & \multicolumn{1}{c}{\sc abs*(bbody+diskbb)}& \multicolumn{1}{c}{\sc abs*(diskbb+bbody)\dag}  \\
\hline
\hline
Model               &   (1)                                          &  (2)                                     &   (3)                           & (4)                            \\
$N_H$              & $17.0\pm0.3$                           & $18.7\pm0.3$                    & $16.9\pm0.3$           & $18.5\pm0.3$             \\
$kT_{BB}$        & $0.54\pm0.01$                         &                                           & $0.56\pm0.01$         & $1.58\pm0.01$           \\
$A_{BB}$          & $1900\pm200$                         &                                           & $1882\pm190$         & $1.6\pm^{+0.7}_{-0.5}$\\
$kT_{DBB}$     &                                                   & $0.66\pm0.02$                  & $1.86\pm0.14$         & $0.68\pm0.01$           \\
$A_{DBB}$       &                                                   & $1170^{+180}_{-150}$      & $0.9^{+0.5}_{-0.3}$  & $1060\pm130$          \\
$\Gamma$        & $3.8\pm0.3$                             & $3.4\pm0.4$                      &                                                                        \\
$A_{\Gamma}$&$6.7^{+2.6}_{-1.9}\times10^{-2}$& $4^{+6}_{-3}\times10^{4}$&                                                                        \\
$\chi^2/dof$      & 1255.2 / 1187                            & 1272.8 / 1187                    & 1246.2 / 1187           & 1262.7 / 1187             \\
\hline
\end{tabular}
\caption{Best fit parameters of the EPIC-pn and MOS2 spectra with two 
component models. The black body normalisation ($A_{BB}$) is in units of 
$(R/km)^2/(D/10~kpc)^2$, where $R$ is the source radius in km units. 
For the definition of the units of all the other parameters, see Tab. \ref{single}. 
\dag These are the parameters associated to the second minimum in the 
$\chi^2$ distribution. }
\label{double}
\end{center}
\end{table*} 

The large residuals at high energy clearly indicate the presence of a 
second emission component. Therefore, we add to the thermal 
emission model the contribution provided by a thermal Comptonisation 
component (reproduced by the {\sc nthComp} model; Zdziarski et al. 1996; 
Zycki et al. 1999). 
Being the high energy cut off outside of the \xmm\ energy band, we fix the 
electron temperature to a high value of $E=50$~keV (consistent 
results are obtained by fixing the electron temperature to e.g. 20 
or 100~keV). The fit with 
an absorbed black body plus thermal Comptonisation model provides 
a satisfactory description of the data (see Tab. \ref{double}). The best 
fit black body temperature is $kT_{BB}=0.54\pm0.01$~keV produced 
from a surface area with radius $R_{BB}=35\pm11$~km. The fit with 
an absorbed multi-temperature disc black body plus Comptonisation 
emission, provides a temperature of $kT=0.66\pm0.02$~keV and 
a very reasonable inner disc radius of $r_{in}\sim27\pm10$~km 
(or $\sim 6.5$~r$_S$, with $r_S=2GM/c^2$, 
where $G$ is the gravitational constant, $M$ is the mass of the primary,
assumed to be $M=1.4$~M$_\odot$ and $c$ is the speed of light), 
however the fit is slightly worse (see Tab. \ref{double}). 
Nevertheless, in both these cases, we observe a rather steep spectral index 
of the Comptonisation component, with values of $\Gamma\sim3.5-3.8$. 
Therefore, we substituted the Comptonisation component with a 
second thermal component. On a statistical ground we obtained an 
improved solution with the sum of an absorbed black body (with 
$kT_{BB}=0.56\pm0.01$~keV) plus a multi-temperature disk black body 
emission 
($kT_{DBB}=1.86\pm0.14$~keV). However, we note that the best fit inner 
radius of the disc black body component is $r_{in}=0.76\pm0.5$~km, 
therefore inside the NS surface (see Tab. \ref{double}). For this reason 
we think that this solution is unlikely. We note a second minimum in the 
$\chi^2$ distribution ($\chi^2=1262.7$ for 1187 dof), leading to a solution 
with $kT_{BB}=1.58\pm0.01$~keV plus a multi-temperature disk black body 
emission 
($kT_{DBB}=0.68\pm0.01$~keV). The best fit inner radius of the disc black 
body component now results to be $r_{in}=26\pm9$~km (corresponding 
to $\sim6$~$r_S$), while the black body component is produced from 
a region with radius of $r\sim1$~km (Tab. \ref{double}). 

In conclusion, the shape of the \xmm\ spectrum is dominated by a thermal 
component (either black body or disc black body, with temperature 
of $kT_{BB}\sim0.55$, $kT_{DBB}\sim0.67$~keV) plus an additional 
component at higher energy that could be reproduced either by a steep 
Comptonization component (perhaps associated with the corona seen in 
accreting systems) or a black body emission (possibly generated by 
the hotter parts on the surface of a NS). 
Any of these possibilities appear viable, based uniquely on the \xmm\ 
data. In \S \ref{secSwift} and \S \ref{secInte}, we will investigate further 
the source radiative process and we will break this model degeneracy 
by studying the \swift\ and \inte\ long term spectral evolution.

\subsection{Ionised absorption}

High inclination accreting BH and NS are known to display ionised absorption 
features, whenever they are observed in the soft state (Neilsen et al. 2009; 
Ponti et al. 2012; 2016). 
No narrow Fe K absorption or emission line is evident in the spectra of \ssv. 
Fitting the spectrum with additional narrow Gaussian absorption lines, 
we compute 
upper limits on the line equivalent width (EW) as stringent as $EW<10$~eV 
to the presence of Fe~{\sc xxv} and Fe~{\sc xxvi} lines, therefore excluding 
the presence of either a disc wind or an ionised atmosphere 
(Diaz-Trigo et al. 2006; Ponti et al. 2012; 2014; 2015; 2016; Miller et al. 2015). 
Similar upper limits are valid for narrow emission lines between 6.4 and 7 keV. 
We also note that the addition of a broad Fe~K emission line is not required. 

\subsection{X-ray properties of the ISM} 

The observation of this bright state of \ssv\ allows us to investigate the 
properties of the ISM along a line of sight placed at less than 
$\sim16$~arcsec from \sgras\ (Baganoff et al. 2016). 
To perform the investigation of the X-ray properties of the ISM, 
we reproduced the source emission with the absorbed disc black body 
plus black body model (model 4 in \S \ref{sec-xmm-meanspec} 
and Tab. \ref{double}). 
The absorption is fitted with the {\sc tbnew} model and the effect of the 
dust scattering halo was taken into account (see Tab. \ref{double}). 
We observed that using the {\sc phabs} absorption model, we measure a 
column density of absorbing material ($N_H=18.3\pm0.3\times10^{22}$
cm$^{-2}$) consistent with the values observed using {\sc tbnew} or {\sc tbabs}. 
On the other hand, a significantly lower value is observed if either 
the {\sc wabs} model ($N_H=12.6\pm0.2\times10^{22}$~cm$^{-2}$) or the 
Anders \& Grevesse (1989) Solar abundances ($N_H=12.1\pm0.2\times10^{22}$
cm$^{-2}$) are used. 

Starting again from the best fit baseline model we explored the possibility 
to constrain the abundances of the various elements. We did so by 
leaving the abundance of any metal producing edges in the 
2-10~keV band free to vary (see Fig. \ref{Spec}). 
We observe that the abundances of 
Si, Cl, Cr, Ca, and Co are unconstrained, therefore we left
their abundances fixed at the Solar value. On the other hand, Iron 
($A_{Fe}=0.5\pm0.4$ Solar) and Argon ($A_{Ar}<0.9$ Solar) are under 
abundant, compared to Solar, while Sulphur 
($A_{S}=1.25^{+0.19}_{-0.14}$ Solar) appears to be slightly overabundant. 
We note that a larger column density of absorbing material is required 
($N_H=20.1\pm1.2\times10^{22}$~cm$^{-2}$), once the Iron abundance 
is left free to vary. 

\section{{\it Long term spectral variations}} 

\subsection{{\it The \swift\ monitoring}} 
\label{secSwift}

\begin{figure}
\centering
\includegraphics[scale=0.47]{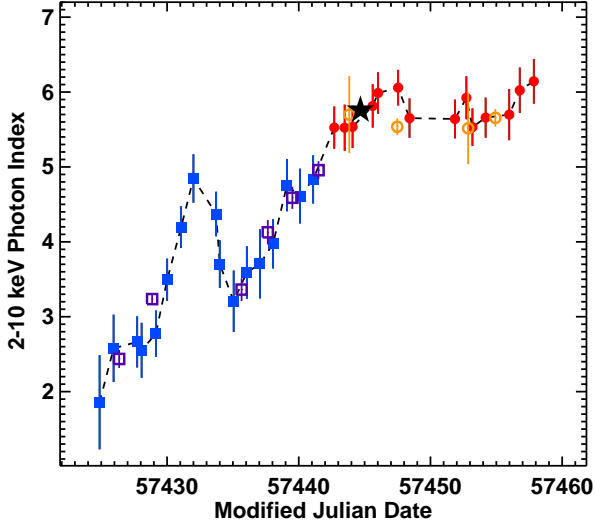}
\caption{Spectral slope evolution of Swift J174540.7-290015. Color symbols 
have the same meaning as in Figure~\ref{fig-tflux}. The photon indices are 
derived by fitting an absorbed power law to the 2-10 keV spectrum of every 
observation, separately, assuming $N_{H} = 25.6\times10^{22}~cm^{-2}$ 
from fitting the \xmm\ spectra. The square symbols 
indicate the hard state observations (filled blue - PC mode, 
empty purple - WT mode), while the circular symbols indicate the soft 
state observations (filled red - PC mode, empty orange - WT mode).
The black star indicates the {\it XMM-Newton} observation.}
\label{fig-specPL}
\end{figure}
\begin{figure}
\centering
\includegraphics[scale=0.47]{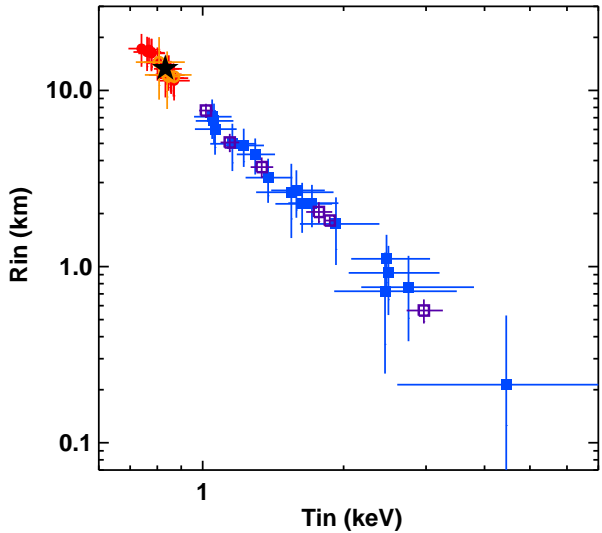}
\caption{The inner disc temperature and flux derived from the best-fit 
disc black body model to the spectrum in every \swift\ observation 
(with $N_H$ fixed at $15.9\times10^{22}~cm^{-2}$). The square symbols 
indicate the hard state observations (filled blue - PC mode, 
empty purple - WT mode), while the circular symbols indicate the soft 
state observations (filled red - PC mode, empty orange - WT mode).
The black star indicates the {\it XMM-Newton} observation.}
\label{fig-tflux}
\end{figure}
\begin{figure}
\includegraphics[width=0.49\textwidth]{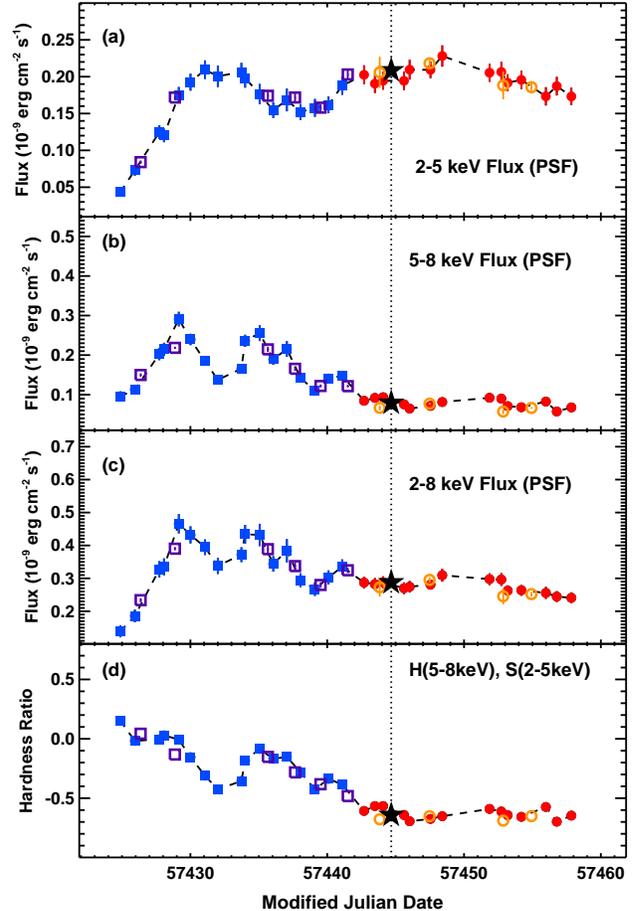}
\caption{Panel-a, b, c are the PSF corrected {\it Swift} XRT flux of 
\ssv\ in the 2-5 keV, 5-8 keV and 2-8 keV band, respectively. 
The blue filled square symbols correspond to the PC mode observations 
when the source spectrum is dominated by a hard X-ray power law 
component. The purple empty squares 
are in the WT mode. The red filled circular symbols correspond to the 
observations when the source spectrum is dominated by a soft X-ray 
thermal component (see Figure~\ref{fig-tflux} and \ref{fig-spec}). 
The orange empty circles are in the WT mode. 
Panel-d shows the hardness ratio between the hard band (5-8 keV) and 
soft band (2-5 keV). In all panels, the black star and the dashed line indicate 
the {\it XMM-Newton} observation on 26-02-2016.}
\label{fig-variability}
\end{figure}
\begin{figure}
\centering
\includegraphics[scale=0.47]{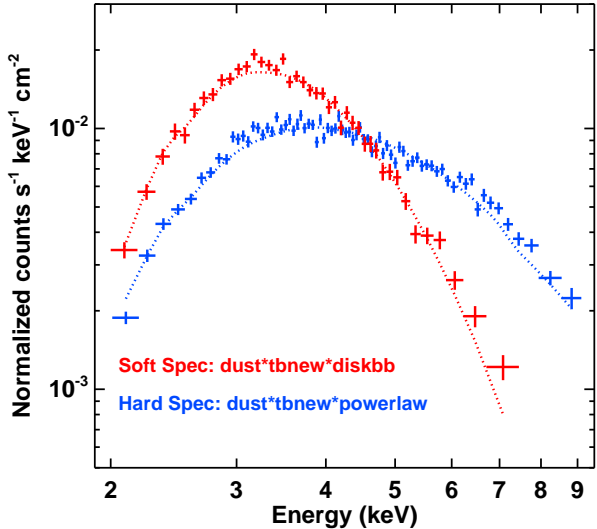}
\caption{Combined, background subtracted, {\it Swift} XRT spectra of the soft 
state (red) and hard state (blue) periods. The combined soft state spectrum is 
well fitted by an absorbed disc black body 
($N_{H} = 16.8\times10^{22}~cm^{-2}$, $kT_{in} = 0.80$ keV), while the hard 
state by an absorbed power law component 
($N_{H} = 17.2\times10^{22}~cm^{-2}$, $\Gamma = 2.56$).}
\label{fig-spec}
\end{figure}

We first fitted all \swift\ spectra with an absorbed power law model 
{\tt dust*tbnew*powerlaw}. Due to the lower statistics of the single $\sim1$~ks 
\swift\ spectra, we assumed a column density of 
$N_H=25.6\times10^{22}cm^{-2}$ (see Tab. \ref{single}) as found in fitting 
the higher resolution {\it XMM-Newton} EPIC spectra. 
This model cannot satisfactorily reproduce all spectra, especially 
when \ssv\ enters softer states. Nevertheless, by applying this  
simple model we can probe the evolution of the spectral shape by 
tracking the variations of the power law photon index.
Figure \ref{fig-specPL} shows the best fit power law photon index 
to the 2-10 keV band. We observe a gradual steepening of the spectral 
shape with time. In particular, very steep spectral shape (with $\Gamma>5$) 
are observed in the later observing period. Such steep spectral indices 
might indicate a thermal nature of the emission mechanism. 
Therefore, we then fitted all the \swift\ spectra with an absorbed 
multi-temperature disk black body model {\tt dust*tbnew*diskbb}. 
Fig.~\ref{fig-tflux} shows the best fit inner disc temperature vs. the 
inner disc radius. A Kendall's rank correlation test provides a correlation 
coefficient of $\rho=-0.95$ and null hypothesis probability of $7\times10^{-25}$. 
Despite its significance, we think that this anti-correlation is induced 
by fitting the data with the wrong spectral model. Indeed, we note that: 
i) the spectra accumulated before MDJ=57442 (blue filled and violet open
squares in Fig.~\ref{fig-specPL} \& \ref{fig-tflux}) are better fit by a power 
law component; ii) the best fit inner disc radii of these spectra is 
significantly smaller than 10~km, implying that the accretion disc 
extends either inside the NS surface or inside the event horizon of 
a stellar mass BH during these observations.
On the other hand, all the {\it Swift} spectra after MJD=57442
(red filled and orange open points in Fig.~\ref{fig-specPL} \& \ref{fig-tflux})
are well fitted by this absorbed thermal emission model.
However, these relatively low S/N spectra cannot rule out the presence
of an additional component in the hard X-ray as observed by {\it XMM-Newton}.
In Fig.~\ref{fig-tflux} the orange open and red filled points cluster around a small 
parameter space, indicating that these spectra also have similar shape.
Moreover, this model cannot reproduces the spectra when there is clearly
a hard component, as represented by the blue and purple squares in Fig.~\ref{fig-tflux}.
Fitting these spectra requires a very high temperature ($kT\sim1-8$~keV) of 
the thermal component and variations of the normalisation that imply 
unlikely variations of the thermal emitting area (see Fig.~\ref{fig-tflux}). 

Finally we adopt model 2 of Tab. \ref{double} that is composed by the sum 
of a thermal multi-temperature disc black body component plus a power 
law {\tt dust*tbnew*(diskbb+powerlaw)}.
This model can well produce all the \swift\ spectra. From top to bottom, 
Fig. \ref{fig-variability} shows the source flux, in the 2-5, 5-8 and 2-8 keV band 
as well as the hardness ratio plotted against the observation date. 
During the first few days of \swift\ observations, \ssv\ showed a clear and 
significant flux increase, by a factor of 3-4. 
Very significant variability, stronger in the hard band, and a hard X-ray
spectrum are observed up to MJD=57442 (see Fig.~\ref{fig-variability}). 
Afterwards, the spectrum softens, with an hardness ratio $HR<-0.5$ 
and no strong variability is observed anymore (see Fig.~\ref{fig-variability}). 

To enhance the signal to noise, we combined\footnote{We use the 
{\sc addspec} tool in the {FTOOLS 6.18} package to derive the combined 
source and background spectra, and group them with {\sc grppha}.
Note that some observations with short exposure and thus very few counts
are excluded from the spectral combination.
These low S/N spectra cannot improve the combined spectra much,
and the inclusion of these spectra would also cause extra empty bins
in the combined spectra due to the spectral combining algorithm.
We also note that the different observations caught \ssv\ at different 
fluxes and hardness ratios, therefore limiting the power of extracting 
detailed spectral information from these averaged spectra. }
the \swift\ spectra obtained before and after MJD=57442.
The combined spectra of these two periods are shown 
in Fig.~\ref{fig-spec} with blue and red symbols, respectively. 
To capture what are the bulk of the spectral variations and 
to avoid model degeneracies (as demonstrated by the higher 
statistic \xmm\ spectrum), we consider here only a single 
emission component model. The spectral degeneracy will be 
broken by the addition of the \inte\ data, above 10~keV.
The hard state spectrum is better fit ($\chi^2=613$ for 538 dof) 
by a simple absorbed power law model with spectral index of $\Gamma=2.56\pm0.07$ 
($N_H=17.2\pm0.5\times10^{22}$~cm$^{-2}$). 
A fit with either a single black body or multi-temperature 
disk black body component provides
a significantly worse fit ($\Delta\chi^2=141$ and 56, respectively).
The combined soft state spectrum is fit by a simple absorbed 
multi-temperature disc black body component with best fit temperature 
$kT_{in}=0.80\pm0.02~keV$  
($N_H=16.8\pm0.7\times10^{22}$~cm$^{-2}$, $\chi^2=374$ for 329 dof)
and inner disc radius of $r_{in}=15\pm2$~km. 
This value is fully compatible with what is expected by an accreting 
neutron star or a black hole. Note that the best fit temperature is slightly lower
than the mean value in Fig.~\ref{fig-tflux}, which is because the column 
density of neutral material ($N_H$) is allowed
to vary when fitting the combined spectra.
For the soft state combined spectrum, a fit with an absorbed power 
law provides a slightly better fit ($\chi^2=353$ for 329 dof),
however the suspiciously large best fit spectral index 
($\Gamma=5.83\pm0.18$, $N_H=26.2\pm1.0\times10^{22}$~cm$^{-2}$)
disfavours this interpretation and leaves the disc black body
model as the more physical interpretation.
The marked spectral and variability difference before and 
after MJD=57442 is reminiscent of spectral transitions in accreting 
X-ray binaries (Fender et al. 2004; Remillard \& McClintock 2006) 
and it suggests that \ssv\ might have undergone a state transition 
during the \swift\ monitoring campaign.  

\begin{figure}
\centering
\includegraphics[scale=0.45]{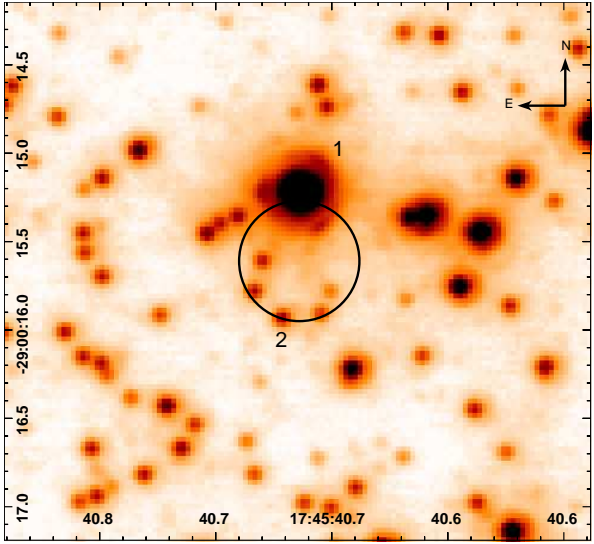}
\caption{VLT-\naco\ archival K-band image of the field of \ssv. 
The \chandra\ error circle (Baganoff et al. 2016) is shown 
by the black circle, with $0.34^{\prime\prime}$ radius. 
The bright candidate counterpart (1) as well as a fainter one (2) 
are indicated. See text for details. North is up, and East is left.}
\label{NACO}
\end{figure}

\subsection{{\it The \inte\ monitoring}}
\label{secInte}

\begin{figure}
\hspace{-0.5cm}
  \includegraphics[width=6.8cm, angle=-90]{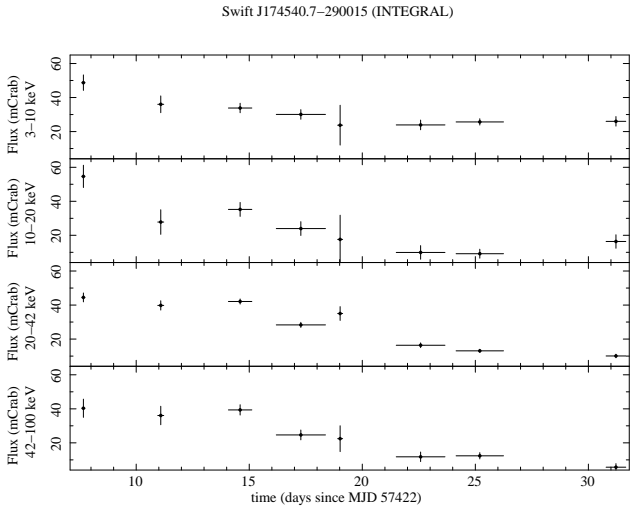}
  \caption{JEM-X and IBIS/ISGRI light-curves of \ssv\ in different 
  energy bands. Each point is integrated for both instruments over 
  an entire \inte\ revolution (from 1643 to 1652). The state transition 
  inferred from the \swift\ data occurred around day 20. The \inte\ 
  data show a simultaneous significant drop of hard X-ray emission 
  ($E>10$~keV), further confirming this interpretation. }
  \label{fig:lcurve}
\end{figure}
To break the degeneracies in the \xmm\ and \swift\ data, induced
by the limited energy band considered, we investigated the source 
behaviour in the \inte\ energy band. 
Figure \ref{fig:lcurve} shows the JEM-X and IBIS/ISGRI light-curves
of \ssv, since the start of the outburst. The source is detected in all 
energy bands. In particular, a significant drop of hard X-ray emission 
($E>10$~keV) is observed after day 20, in correspondence to the 
state transition inferred from the \swift\ data (Fig. \ref{fig:lcurve}),
therefore further confirming this scenario (see \S \ref{secSwift}).

We accumulated all \inte\ data that are quasi-simultaneous 
to \swift\ observations and we created a combined "hard state" 
and "soft state" spectrum (summing all ISGRI and JEM-X spectra 
from revolution 1643-1648 and 1649-1652, corresponding to 
MJD 57429-57441 and 57443-57453, respectively). We then 
performed a broad-band fit by considering these \inte\ and 
\swift\ spectra. As already clear 
from the analysis of the \xmm\ and \swift\ data, a single component 
emission model is not an adequate description of the emission 
of \ssv\ (see \S \ref{sec-xmm-meanspec} and \S \ref{secSwift}). 
Therefore, we began by applying the best fit double component models 
fitting the \xmm\ spectrum (see Tab. 3) to the soft state 
cumulative \swift\ and \inte\ spectra. In all models that 
we explored, we applied an inter-normalisation factor 
(see Tab. \ref{integr}) between the different instruments.

We first considered the absorbed double thermal component 
({\sc cons*abs*(diskbb+bbody)}, see model 4 in Tab. \ref{double}). 
This model (see \S \ref{sec-xmm-meanspec}
and Tab. \ref{double}), being composed by two thermal 
components peaking below 2~keV, predicts very little emission 
above $\sim20$ keV. On the contrary, even in the soft state, \ssv\ emits
significant radiation above 20~keV. To try to reproduce the 
hard X-ray emission, this model increases the best fit temperature 
of the disc black body component and lowers the disc inner radius
to unrealistic values. Indeed, the best fit temperature is 
$kT_{DBB}=10^{+2.6}_{-1.6}$~keV and the inner disc radius is 
$r_{in}\sim0.025$~km, the latter being too small for either a NS or a BH. 
For this reason we rule out this model. 

We then tested the absorbed black body plus Comptonisation 
component ({\sc cons*abs*(bbody+nthComp)}, model 1 in in Tab. 
\ref{double}) as well as the absorbed disc black body plus 
Comptonisation component ({\sc cons*abs*(diskbb+nthComp)}
model 2 in Tab. \ref{double}). Incidentally, we note that even adding 
the \inte\ data, the energy of the cutoff in the electron distribution is 
unconstrained ($E>15$~keV). Therefore we fix this value to 50~keV,
as in the previous fits (\S \ref{sec-xmm-meanspec} and Tab. \ref{double}). 
Both these thermal plus Comptonisation models provide a good 
description of the data ($\chi^2=357.0$ for 331 dof, $\chi^2=355.9$ for 
331 dof, respectively). We note that the best fit parameters of the 
absorber, black body and disc black body components are consistent 
with the ones obtained by fitting the higher statistics \xmm\ spectrum 
alone (see Tab. \ref{double}). 
However, the addition of the \inte\ data allows us to obtain 
now more reliable constraints on the parameters of the Comptonisation 
component. In particular, the photon index of the Comptonisation 
component ($\Gamma=2.1$) is now consistent with the values typically 
observed in accreting compact objects. Moreover, the detection of X-ray 
emission up to $\sim50$~keV, allowed us to rule out that the soft state 
emission is only composed by a combination of pure thermal components. 

\begin{figure*}
\hspace{-0.5cm}
  \includegraphics[width=7.1cm, angle=-90]{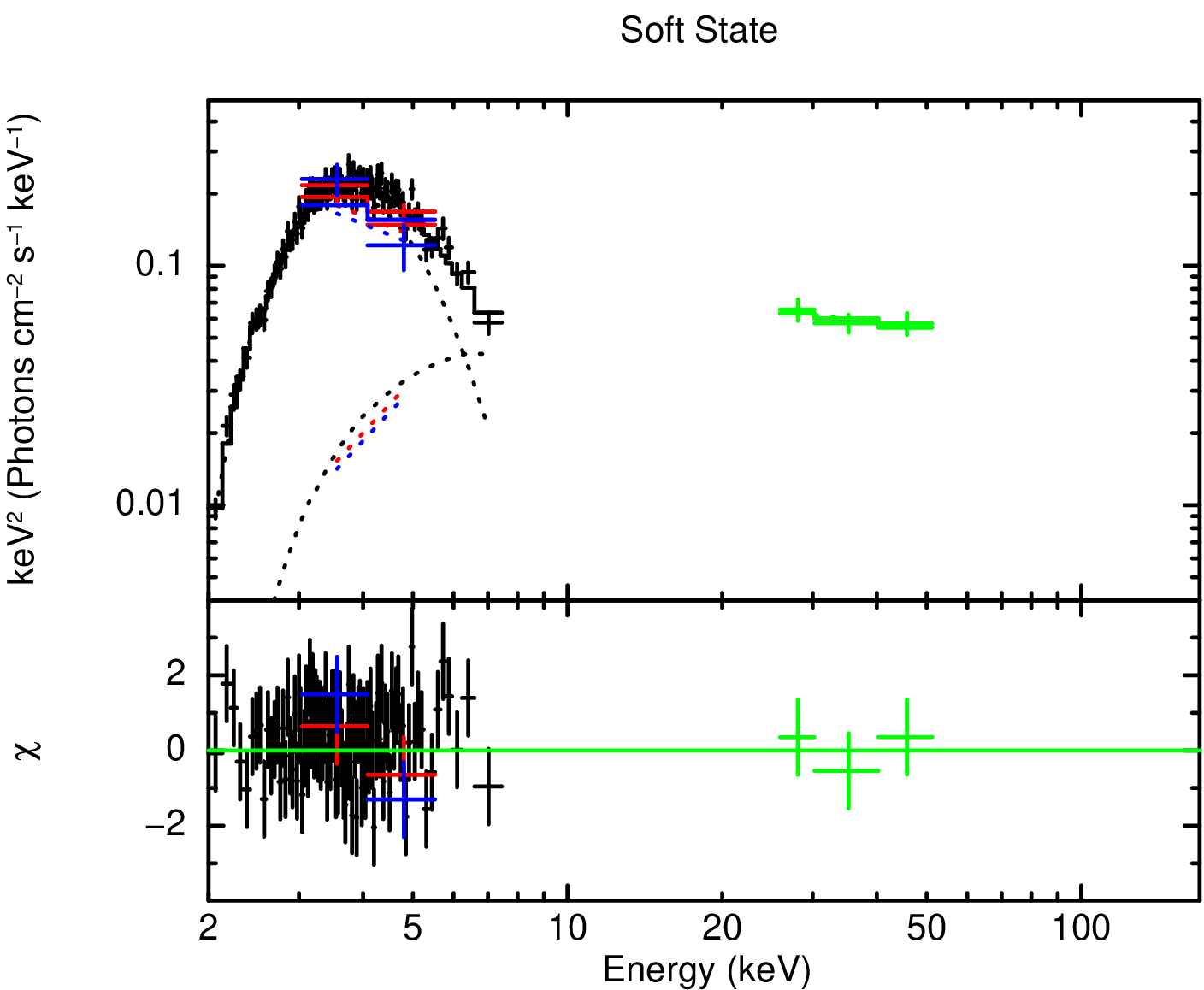}
  \includegraphics[width=7.1cm, angle=-90]{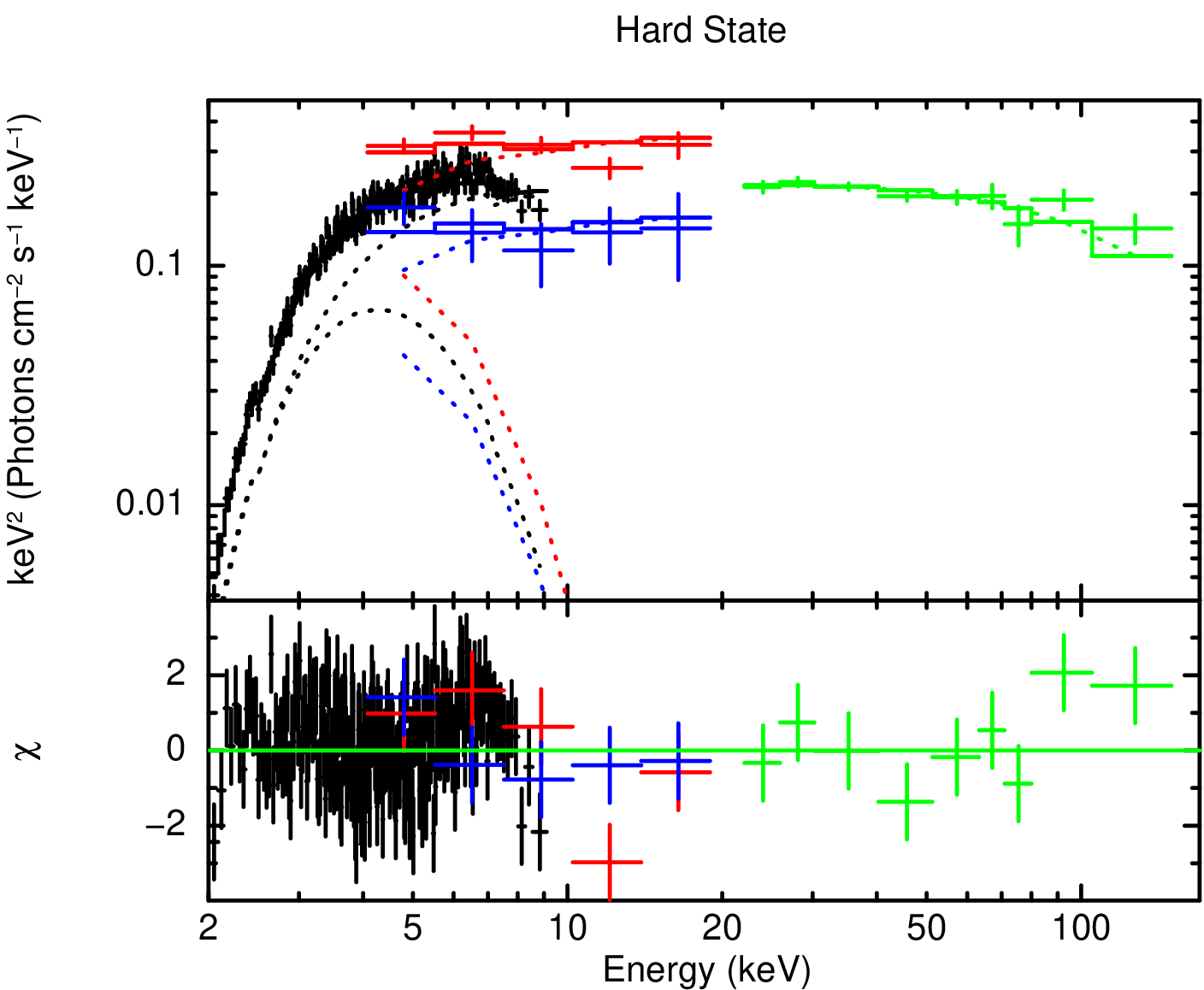}
  \caption{The broad-band spectra of \ssv\ extracted during the soft 
  and hard states (as defined in \S~\ref{secInte}). For the hard state 
  {\it (right panel)} we used data from the \inte\ revolutions 1643-1648 
  (in black, red, blue and green are the \swift\ XRT, JEM-X1, JEM-X2 
  and IBIS/ISGRI). For the soft state {\it (left panel)}, data from 
  revolutions 1649-1652 have been used (same color scheme as before). 
  The soft state spectrum 
  is dominated by an absorbed thermal component below $\sim10$~keV
  and a Comptonisation tail at higher energies. The hard state spectrum 
  is dominated by an absorbed Comptonisation component inducing 
  significantly larger high energy flux. See text for details.}
  \label{fig:spectra}
\end{figure*}

We then also fitted the combined hard state spectrum (see Fig. \ref{fig:spectra}). 
The broad band spectra can be fit by an absorbed Comptonisation 
component plus either a black body or a disc black body (see models 1H and 
2H in Tab. \ref{integr}). 
None of these two models provide a completely acceptable fit
($\chi^2=659.2$ for 552 dof and $\chi^2=648.9$ for 552 dof, respectively). 
We attribute this to the strong spectral and flux evolution (e.g., variation in 
photon index) observed during the hard state (see \S \ref{secSwift}), 
that can not be properly taken into account by the fit of these combined 
spectra. Therefore, we refrain the reader from extracting too detailed 
information from these spectra\footnote{For example, we note that 
some inter-normalisation factors are not consistent with 1 and that 
significant residuals are still present.}. It is clear, however, that 
the hardness ratio changed by a factor of $\sim3.4$ between the hard 
($F_{H_{10-100~keV}}=6.5\times10^{-10}$ erg cm$^{-2}$ s$^{-1}$, 
$F_{H_{2-10~keV}}=3.8\times10^{-10}$ erg cm$^{-2}$ s$^{-1}$) and 
soft ($F_{S_{10-100~keV}}=1.4\times10^{-10}$ erg cm$^{-2}$ s$^{-1}$,
$F_{S_{2-10~keV}}=2.8\times10^{-10}$ erg cm$^{-2}$ s$^{-1}$) state 
(see Fig. \ref{fig:spectra}). This reinforces the interpretation that the source 
underwent a state change during the monitored period. 

\begin{table*} 
\begin{center}
\begin{tabular}{ c c c c c c}
\hline
\hline
\multicolumn{6}{c}{Broad band fit of composite soft and hard state spectra~{\sc abs = dust*tbnew}}\\
& \multicolumn{3}{c}{Composite Soft State} & \multicolumn{2}{c}{Composite Hard State}\\
& \multicolumn{1}{c}{\sc abs*(diskbb+bbody)}  & \multicolumn{1}{c}{\sc abs*(bbody+nthComp)} & \multicolumn{1}{c}{\sc abs*(diskbb+nthComp)} & \multicolumn{1}{c}{\sc abs*(diskbb+nthComp)} & \multicolumn{1}{c}{\sc abs*(bbody+nthComp)} \\
\hline
\hline
Model               &   (4S)                                          &  (1S)                                     &   (2S)                           &  (2H)                                     &   (1H)                     \\
$N_H$              & $15.9\pm0.9$                           & $17.1\pm1.1$                    & $18.8\pm1.0$            & $14.3\pm1.0$                    & $13.3\pm1.0$       \\
$kT_{BB}$        & $0.62\pm0.02$                         & $0.57\pm0.03$                   &                                  &                                           & $0.79\pm0.1$       \\
$A_{BB}$          & $1100\pm350$                         &  $1800\pm800$                 &                                   &                                           & $68_{-23}^{+63}$ \\
$kT_{DBB}$     & $10^{2.6}_{-1.6}$                      &                                           & $0.67\pm0.04$          & $1.16\pm0.18$                  &                              \\
$A_{DBB}$       &$1_{-0.5}^{+1.2}\times10^{-3}$  &                                           & $1100_{-400}^{+700}$& $14_{-8}^{+21}$              &                               \\
$\Gamma$        &                                                  & $2.1\pm0.4$                      & $2.1\pm0.4$              & $1.92\pm0.05$                  & $1.94\pm0.06$    \\
$A_{\Gamma}$&                                          &$1.4_{-0.5}^{+1.0}\times10^{-2}$& $2.9\pm1.7\times10^{-2}$&$8.7\pm5\times10^{-3}$& $2.4\pm1\times10^{-3}$ \\
$C_{JEMX1}$   & $<1.16$                                   & $0.96\pm0.2$                     & $0.96\pm0.2$             & $1.5\pm0.2$                     & $1.5\pm0.1$      \\
$C_{JEMX2}$   & $<1.04$                                   & $1.3_{-0.7}^{+1.5}$           & $1.4_{-0.7}^{+1.6}$     & $0.98_{-0.13}^{+0.18}$     & $0.92\pm0.12$  \\
$C_{ISGRI}$     & $<1.06$                                   & $0.89\pm0.18$                  & $0.9\pm0.2$                & $0.85\pm0.12$                  & $0.70\pm0.12$   \\
$\chi^2/dof$      & 369.5 / 331                              & 357.0 / 331                        & 355.9 / 331                  & 648.9 / 552                       & 659.2 / 552    \\
\hline
\hline
\end{tabular}
\caption{Best fit parameters of the combined, nearly simultaneous \inte\ 
and \swift\ soft state and hard state spectra of \ssv. The same models applied 
to the higher statistics \xmm\ spectrum have been fitted to the combined 
broad band soft state (1S, 2S and 4S) spectrum, as well as to the hard 
state (1H, 2H) one. For the definition of 
the units of all the other parameters, see Tab. \ref{single} and \ref{double}. }
\label{integr}
\end{center}
\end{table*}

\section{Possible near infra-red counterpart}

The location of \ssv\ was observed with the Gamma-Ray burst Optical
Near-infrared Detector (GROND; Greiner et al. 2008) mounted at the
MPG/ESO 2.2m telescope in La Silla, on 2016 Febuary 19.35 UT.
Observations were obtained simultaneously in the J, and H photometric
bands for 640~ s integration time each\footnote{GROND observes
simultaneously in g$^\prime$, r$^\prime$, i$^\prime$, z$^\prime$, J, H, K.
Due to the extreme Galactic foreground reddening affecting the optical bands
and saturation affecting the K-band, only the J and H band data are discussed 
here.}. The data were reduced with the standard tools and methods as described
in Kr\"uhler et al. (2008). The photometry was measured from circular
apertures with radii corresponding to the image full width at half maximum
($1.1^{\prime\prime}$ in J and $1.4^{\prime\prime}$ in H) and calibrated against 
2MASS stars in the same field.

A single bright source, reported also by Masetti et al. (2016) from archival 
VVV imaging, is seen at the \chandra\ X-ray position. The apparent aperture 
magnitudes (in the Vega system) are $J=15.97\pm0.12$ and 
$H=12.76\pm0.13$. The location of the new transient was also contained 
in a series of 15 GROND observations obtained on 2015 April 26.35-26.42 UT 
when monitoring another Galactic Center transient. No variability was found 
for the bright near-IR source in the J or H band during this set of observations 
and the median magnitudes of $J=16.10\pm0.11$ (seeing $1.1^{\prime\prime}$) 
and $H=12.57\pm0.10$ ($1.6^{\prime\prime}$) are consistent within errors 
with the post-outburst measurements.

The near-IR source density at the position of the transient is very high and
GROND photometry corresponds with the co-addition of the flux of all sources 
within the aperture. Therefore, we can only conclude, that the brightest 
source within the \chandra\ error circle, dominating the total emission, 
remained constant between the pre- and post-outburst observations and 
none of the fainter objects brightened to a level where it would increase 
the overall emission within the aperture by more than 0.1 magnitude.

The GC is routinely observed with adaptive optics using \naco\ 
at the VLT (see Gillessen et al. 2009 for a description of the program and 
data reduction). Figure \ref{NACO} shows a stacked \naco\ K band image from
archival (pre-outburst) observations. 
Besides the bright 
source dominating the GROND photometry, multiple fainter objects are 
resolved either inside or in the close proximity to the \chandra\ error 
region (black circle). 
For the bright source (see Fig. \ref{NACO}) we measure magnitudes of 
$H=12.98\pm0.06$ and $K=10.81\pm0.03$ while, for comparison, the 
fainter star (indicated with 2 in Fig. \ref{NACO}) has 
$H_{star2} = 17.24 \pm 0.06$ and $K_{star2} = 15.23 \pm 0.03$. 
The MPE-IR group obtained new \naco\ data on 2016 March 16. 
No significant variation in brightness of any of the sources within the 
\chandra\ error circle is observed (Gillessen private communication).

The probability of having such a bright star within the \chandra\ error circle 
of $\sim0.4$ arcsec$^2$ can be computed from number of stars 
with $m_{Ks}>11$~mag within an annulus centred on \sgras\ and with inner 
and outer radii of $16-5$~arcsec and $16+5$~arcsec, respectively. 
This annulus includes the position of the putative near-IR counterpart. 
The observed surface density within the annulus is 0.032 stars arcsec$^{-2}$ 
implying a chance coincidence probability of $\sim1$~\% of finding a source 
with $m_{Ks}>11$~mag within the \chandra\ error circle, i.e. that the bright 
source is not associated with the the X-ray transient.

Using the average reddening towards the Galactic Center of 
$A_{\rm K}=2.4\pm0.1$ mag (Fritz et al. 2011), the absolute 
K-band magnitude is $M_{\rm K}=-6.1$. The star spectroscopically shows 
the CO band heads in the K-band, which excludes stars hotter than spectral 
class F (Wallace \& Hinkle 1997). The absolute magnitude would suggest 
it to be a late type giant or super giant.

\section{Discussion}

We report the identification of a new accreting X-ray binary, 
\ssv\ and results from the early phases of its X-ray emission evolution.
\ssv\ reached a peak flux at an absorbed value of 
$F=6\times10^{-10}$ erg cm$^{-2}$ s$^{-1}$ in the 2-10 keV 
band. During the \xmm\ observation, \ssv\ shows an absorbed 
and un-absorbed 2-10~keV flux of $F_{abs}=2.9\times10^{-10}$ 
and $F_{un-abs}=14.5\times10^{-10}$ erg cm$^{-2}$ s$^{-1}$, 
respectively, that corresponds to $F_{un-abs}=52.8\times10^{-10}$ erg 
cm$^{-2}$ s$^{-1}$ in the 0.1-10~keV band. 
During the \xmm\ observation, \ssv\ showed 
an observed 2-10 keV luminosity of $L_{abs}=2.1\times10^{36}$
erg~s$^{-1}$ and an un-absorbed luminosity of 
$L_{un-abs}=3.8\times10^{37}$ erg s$^{-1}$ in the 0.1-10~keV 
band. Therefore, if \ssv\ is a NS with $M_{NS}=1.4 M_{\odot}$, 
the observed luminosity is $L\sim20$~\% of the Eddington limit 
during the \xmm\ observation ($L<20$~\% if \ssv\ is a BH).
Given these luminosities, \ssv\ does not belong to the class of faint 
X-ray transients (Wijnands et al. 2006). 

We point out that \ssv\ is located at less than 16 arcsec from \sgras\ 
(the supermassive BH at the GC), less than 17 arcsec from \sgr\ (Rea et al. 2013; 
Mori et al. 2013; Coti-Zelati et al. 2015) and  less than 1.7 arcmin from 
the neutron star low mass X-ray binary AX~J1745.6-2901 (Ponti et al. 2015, 2016), 
therefore it lies in the field of view of all \xmm\ observations pointed 
to one of these sources (Ponti et al. 2015a,b,c; 2016). 
No evidence for previous outbursts of \ssv\ is present in the \xmm, 
\chandra\ and \swift\ archive (Degenaar et al. 2012; Ponti et al. 2015a,b). 

We compute the upper limit on the X-ray emission of \ssv\ during 
quiescence by stacking all \chandra\ ACIS-I observations between 1999 
and 2011, for a total exposure of $\sim1.8$~Ms. We observe 54 counts 
within $0.34^{\prime\prime}$ from the position of \ssv, in the 0.5-10~keV band. 
This translates in a $95$~\% upper limit of $3.5\times10^{-5}$ cts 
s$^{-1}$ (Gehrels 1986). Assuming the quiescent spectrum of \ssv\ 
is described either by an absorbed ($N_H=18.7\times10^{22}$ cm$^{-2}$)
power law (with spectral index $\Gamma=2$) or a black body 
($kT_{BB}=0.3$~keV) this can be translated into an un-absorbed 
luminosity of $L_{\rm PL_{0.5-10}}<5\times10^{31}$ erg s$^{-1}$ 
or  $L_{\rm BB_{0.5-10}}<1.7\times10^{33}$ erg s$^{-1}$. 
This upper limit on the quiescent X-ray luminosity is consistent 
with both a NS or BH origin of \ssv\ (Narayan et al. 1997; Garcia et al. 
2001; Rea et al. 2011; Armas Padilla et al. 2014).

Radio observations did not detect either a steady or periodic counterpart 
to \ssv. Steady emission is expected under two scenarios. First,
the fundamental plane of BH activity can be used to predict the radio
luminosity under the assumption that the system is in the low hard state 
(Plotkin et al. 2012). For a 10 $M_{BH}$ black hole with the unabsorbed 
X-ray luminosity observed by \xmm, we predict a radio flux density of 6~mJy. 
The scatter in the fundamental plane is approximately an order of magnitude 
so that values as high as 100~mJy are possible. The imaging non-detection 
with an rms of 6~mJy, thus, can neither rule out nor confirm the presence of a jet
in this system. We note that by the time the first VLA observation has been 
taken, \ssv\ already transitioned to the soft state. Second, bright radio emission 
can be produced through interaction of a jet with the dense interstellar 
medium of the Galactic center. 
The low-mass X-ray binary CXOGC J174540.0-290031, located only 0.1 pc 
in projection from \sgras, reached a peak flux density of 100~mJy that was 
resolved into two lobes on either sides of the X-ray (Bower et al. 2005). 
The proximity of \ssv\ to Sgr~A~West suggests that it could produce similar 
interactions but we find no evidence for these.

The emission of \ssv, during the \xmm\ observation is 
dominated by a thermal component that can be well fitted either by 
multi-temperature disc black body, or black body emission. 
In addition, emission at higher energies (up to 20-50~keV) is clearly 
detected by \inte. The high energy emission is a clear sign of a 
Comptonization component, ruling out a double thermal component
consistent with the fits limited to energies below 10~keV.
The disc black-body plus Comptonisation scenario is perfectly 
compatible either with a NS or BH primary.
In this framework, the Comptonization component might be associated 
with the coronal emission typically observed in accreting sources and the 
multi-temperature disc black body could be produced by a standard 
accretion disc extending down to $\sim6.5$~r$_S$ for a neutron star 
($M_{NS}=1.4$~M$_\odot$) or $\sim2$~r$_S$ for a BH with 
$M_{BH}=5$~M$_\odot$. 
In the second scenario (model 1 and 1S in Tab. \ref{double} and 
\ref{integr}, respectively), the presence of the black body emission 
suggests that the primary is a NS. Also in this scenario, the Comptonisation 
might be due to coronal emission, but the soft radiation would be dominated 
by the boundary layer emission.
However, the best fit black body radius appears to be too large 
($R_{BB}=35\pm11$~km) to be associated with a conceivable NS. 
Nevertheless, we can not exclude that the large observed area, attributed 
to the black body emission, might be induced by the contribution from 
a disc emission component not considered in our modelling. 
Alternatively, the black body component might be in part produced 
by hot spots on the surface of the NS. Would this indeed be the case, 
then a pulsation should be present.
However, this pulsation might be undetected either because of the 
weakness of the black body emission or because \ssv\ is a fast 
spinning millisecond pulsar (e.g., faster than our timing resolution, or a 
weak signal possibly also smeared by a Doppler orbital shift).

From the spectral behaviour of \ssv\ we can rule out a high mass 
X-ray binary (HMXB) nature of the source. The spectra of HMXBs 
are characterised by a dominant hard power law in the 0.2$-$10.0 keV 
band with typical photon indices of 1.0$-$1.5 for supergiant systems 
and around 0.9 for Be/X-ray binaries (e.g. Haberl et al. 2008; 
Bozzo et al. 2012; Bodaghee et al. 2012; Walter et al. 2015). 
HMXBs also do not show spectra state transitions as observed from \ssv. 

\ssv\, was initially identified as a possible magnetar due to its fast 
outburst rise. However, many observational characteristics of this 
new transient are at variance with what is usually observed in magnetar 
outbursts (see Rea \& Esposito 2011 and Turolla, Zane \& Watts 2015 
for a review on outbursts and on magnetars in general, respectively). 
The strong variability of the outburst flux decay (see Fig.\ref{fig-variability}) 
is unseen in the outburst decay of magnetars, which are usually observed 
to decay smoothly toward quiescence. This is also in line with the 
interpretation of these outbursts as being due to a large energy injection 
in the magnetar crust triggered by the instability of the internal or 
external magnetic field (possibly by a strongly twisted magnetic bundle), 
with a consequent crustal heating and subsequent cooling 
(Beloborodov 2009; Pons \& Rea 2012). Furthermore, if placed 
at 8\,kpc the peak luminosity of the outburst is too high with respect 
to what is observed for magnetars, $L_{X}<10^{36}$\,erg/s, which 
is a limit believed to be regulated by the strong dependence 
on temperature of the crustal neutrino emissivity (Pons \& Rea 2012). 
The deep upper limits on the PF of this new transient (see Fig.\,\ref{PF}) 
are rather unlikely (although a few exceptions exists) compared 
with the usual strength ($\sim$20-80\%) of the periodic 
signals (around 2-10\,s) observed for magnetars. 
Therefore, we believe the magnetar interpretation for this new transient 
is extremely unlikely.

Observations of the X-ray reflection nebulae (irradiated molecular 
clouds within a few hundred parsecs from \sgras) suggest an either 
Solar or supersolar Iron abundance of the GC ISM 
(Revnitsev et al. 2004; Terrier et al. 2010; Ponti et al. 2010; 2013; 2014b;
Zhang et al. 2015). 
The observed low Iron abundance ($A_{Fe}=0.5\pm0.4$ Solar) of 
the neutral absorbing material towards \ssv\ is therefore rather surprising 
also with respect to the accepted iron gradient in the Galactic disk 
(Friel et al. 2002; Pedicelli et al. 2009). We note that the measured value is consistent 
with the Iron abundance towards 4U~1820-303, an X-ray binary located 
at a distance of $\sim7.6$~kpc, at less than 1~kpc from the GC (towards 
$l=2.79^\circ$ and $b=-7.91^\circ$; Kuulkers et al. 2003; Pinto et al. 2010). 
The observed under abundance is also in line with the values measured 
towards a small sample of bright Galactic X-ray binaries (Juett et al. 2006). 
We attribute this to depletion of Iron into dust grains in the ISM (see Juett 
et al. 2006). Indeed, depletion of elements in dust grains can reduce the 
effective cross section for absorption. In grains, the optical depth can be 
higher than one, therefore absorption will occur primarily on the surface. 
Moreover, Iron atoms located deeper into the grain will be shielded and they 
will not contribute to the absorption, effectively reducing the strength 
of the element absorption edge (e.g. Juett et al. 2006). 
The similarity of the observed Iron abundance to the ones measured 
along many other line of sights, is consistent with the idea that 
a significant fraction of the absorption is produced in the ISM along 
the Galactic arms and therefore it does not carry information about 
the metal abundances characteristic of the GC region. 
Future detailed modelling of the dust scattering halo of \axj\ and \ssv\ 
will better constrain this. 

Given the long-term X-ray variability and spectral decomposition, 
we then identify \ssv\ as an accreting X-ray binary hosting a neutron 
star or a black-hole, most likely with a low mass star companion.
The discovery and identification of this new GC transient as well as 
the continuous X-ray monitoring of the central parsec is improving 
the determination of the population of stellar remnants orbiting \sgras. 

\section*{Acknowledgments}

The authors wish to thank N. Schartel and the \xmm\ team for performing 
the \xmm\ observation and the "MPE IR-group Galactic Center team" 
for sharing the reduced, cleaned, calibrated \naco\ images and photometry 
as well as for help with the interpretation and discussion. 
The GC \xmm\ monitoring project is supported by the Bundesministerium 
f\"{u}r Wirtschaft und Technologie/Deutsches Zentrum f\"{u}r Luft- und 
Raumfahrt (BMWI/DLR, FKZ 50 OR 1408) and the Max Planck Society. 
The National Radio Astronomy Observatory is a facility of the 
National Science Foundation operated under cooperative agreement 
by Associated Universities, Inc.
NR and FCZ acknowledge financial support from a Dutch NWO Vidi grant. 
NR also acknowledges support from Spanish grants AYA2015-71042 and
SGR2014-1073
This research has made use primarily of data obtained with \xmm, an ESA 
science mission with instruments and contributions directly funded by ESA 
Member States and NASA, and on data obtained from the \swift\ Data Archive.
Part of the funding for GROND (both hardware as well as personnel) was 
generously granted from the Leibniz-Prize to Prof. G. Hasinger 
(DFG grant HA 1850/28-1). We thank P. Wiseman, J. Greiner (both MPE) 
and A. Hempel (PUC) for their support with the GROND observation.


\begin{thebibliography}{1}

\bibitem[Anders \& Grevesse(1989)]{1989GeCoA..53..197A} Anders, E., \& Grevesse, N.\ 1989, \gca, 53, 197 

\bibitem[Armas Padilla et al.(2014)]{2014MNRAS.444..902A} Armas Padilla, M., Wijnands, R., Degenaar, N., et al.\ 2014, \mnras, 444, 902 

\bibitem[Baganoff et al.(2016)]{2016ATel.8746....1B} Baganoff, F.~K., Corrales, L.~R., Neilsen, J., et al.\ 2016, The Astronomer's Telegram, 8746,  

\bibitem[Beloborodov(2009)]{2009ApJ...703.1044B} Beloborodov, A.~M.\ 2009, \apj, 703, 1044 

\bibitem[Bodaghee et al.(2012)]{2012AIPC.1427...52B} Bodaghee, A., Tomsick, J.~A., Rodriguez, J., et al.\ 2012, American Institute of Physics Conference Series, 1427, 52 

\bibitem[Bozzo et al.(2012)]{2012A&A...544A.118B} Bozzo, E., Pavan, L., Ferrigno, C., et al.\ 2012, \aap, 544, A118 

\bibitem[Bozzo et al.(2016)]{2016A&A...589A..42B} Bozzo, E., Pjanka, P., Romano, P., et al.\ 2016, \aap, 589, A42 

\bibitem[Bower et al.(2005)]{2005ApJ...633..218B} Bower, G.~C., Roberts, D.~A., Yusef-Zadeh, F., et al.\ 2005, \apj, 633, 218 

\bibitem[Bower et al.(2016)]{2016ATel.8793....1B} Bower, G.~C., Demorest, P., Baganoff, F., et al.\ 2016, The Astronomer's Telegram, 8793,  

\bibitem[Campana(2009)]{2009ApJ...699.1144C} Campana, S.\ 2009, \apj, 699, 1144 

\bibitem[Courvoisier et al.(2003)]{2003A&A...411L..53C} Courvoisier, T.~J.-L., Walter, R., Beckmann, V., et al.\ 2003, \aap, 411, L53 

\bibitem[Coti Zelati et al.(2015)]{2015MNRAS.449.2685C} Coti Zelati, F., Rea, N., Papitto, A., et al.\ 2015, \mnras, 449, 2685 

\bibitem[Degenaar et al.(2015)]{2015JHEAp...7..137D} Degenaar, N., Wijnands, R., Miller, J.~M., et al.\ 2015, Journal of High Energy Astrophysics, 7, 137 

\bibitem[Degenaar et al.(2012)]{2012A&A...545A..49D} Degenaar, N., Wijnands, R., Cackett, E.~M., et al.\ 2012, \aap, 545, A49 

\bibitem[De Marco et al.(2015)]{2015MNRAS.454.2360D} De Marco, B., Ponti, G., Mu{\~n}oz-Darias, T., \& Nandra, K.\ 2015, \mnras, 454, 2360 

\bibitem[Dexter \& O'Leary(2014)]{2014ApJ...783L...7D} Dexter, J., \& O'Leary, R.~M.\ 2014, \apjl, 783, L7 

\bibitem[D{\'{\i}}az Trigo et al.(2006)]{2006A&A...445..179D} D{\'{\i}}az Trigo, M., Parmar, A.~N., Boirin, L., M{\'e}ndez, M., \& Kaastra, J.~S.\ 2006, \aap, 445, 179 

\bibitem[Eatough et al.(2013)]{2013Natur.501..391E} Eatough, R.~P., Falcke, H., Karuppusamy, R., et al.\ 2013, \nat, 501, 391 

\bibitem[Esposito et al.(2016)]{2016ATel.8684....1E} Esposito, V., Kuulkers, E., Bazzano, A., et al.\ 2016, The Astronomer's Telegram, 8684,  

\bibitem[Faucher-Gigu{\`e}re \& Loeb(2011)]{2011MNRAS.415.3951F} Faucher-Gigu{\`e}re, C.-A., \& Loeb, A.\ 2011, \mnras, 415, 3951 

\bibitem[Fender \& Belloni(2012)]{2012Sci...337..540F} Fender, R., \& Belloni, T.\ 2012, Science, 337, 540 

\bibitem[Fender et al.(2004)]{2004MNRAS.355.1105F} Fender, R.~P., Belloni, T.~M., \& Gallo, E.\ 2004, \mnras, 355, 1105 

\bibitem[Frank et al.(1987)]{1987A&A...178..137F} Frank, J., King, A.~R., \& Lasota, J.-P.\ 1987, \aap, 178, 137 

\bibitem[Friel et al.(2002)]{2002AJ....124.2693F} Friel, E.~D., Janes, K.~A., Tavarez, M., et al.\ 2002, \aj, 124, 2693 

\bibitem[Garcia et al.(2001)]{2001ApJ...553L..47G} Garcia, M.~R., McClintock, J.~E., Narayan, R., et al.\ 2001, \apjl, 553, L47 

\bibitem[Gehrels(1986)]{1986ApJ...303..336G} Gehrels, N.\ 1986, \apj, 303, 336 

\bibitem[Genzel et al.(2010)]{2010MNRAS.407.2091G} Genzel, R., Tacconi, L.~J., Gracia-Carpio, J., et al.\ 2010, \mnras, 407, 2091 

\bibitem[Gillessen et al.(2013)]{2013IAUS..289...29G} Gillessen, S., Eisenhauer, F., Fritz, T.~K., et al.\ 2013, Advancing the Physics of Cosmic Distances, 289, 29 

\bibitem[Greiner et al.(2008)]{2008PASP..120..405G} Greiner, J., Bornemann, W., Clemens, C., et al.\ 2008, \pasp, 120, 405 

\bibitem[Haberl et al.(2008)]{2008A&A...489..327H} Haberl, F., Eger, P., \& Pietsch, W.\ 2008, \aap, 489, 327 

\bibitem[Hopman(2009)]{2009ApJ...700.1933H} Hopman, C.\ 2009, \apj, 700, 1933 

\bibitem[Israel \& Stella(1996)]{1996ApJ...468..369I} Israel, G.~L., \& Stella, L.\ 1996, \apj, 468, 369 

\bibitem[Jansen et al.(2001)]{2001A&A...365L...1J} Jansen, F., Lumb, D., Altieri, B., et al.\ 2001, \aap, 365, L1 

\bibitem[Juett et al.(2006)]{2006ApJ...648.1066J} Juett, A.~M., Schulz, N.~S., Chakrabarty, D., \& Gorczyca, T.~W.\ 2006, \apj, 648, 1066 

\bibitem[Kaspi et al.(2014)]{2014ApJ...786...84K} Kaspi, V.~M., Archibald, R.~F., Bhalerao, V., et al.\ 2014, \apj, 786, 84 

\bibitem[Kr{\"u}hler et al.(2008)]{2008ApJ...685..376K} Kr{\"u}hler, T., K{\"u}pc{\"u} Yolda{\c s}, A., Greiner, J., et al.\ 2008, \apj, 685, 376-383 

\bibitem[Kuulkers et al.(2003)]{2003A&A...399..663K} Kuulkers, E., den Hartog, P.~R., in't Zand, J.~J.~M., et al.\ 2003, \aap, 399, 663 

\bibitem[Lebrun et al.(2003)]{2003A&A...411L.141L} Lebrun, F., Leray, J.~P., Lavocat, P., et al.\ 2003, \aap, 411, L141 

\bibitem[Lee(1995)]{1995MNRAS.272..605L} Lee, H.~M.\ 1995, \mnras, 272, 605 

\bibitem[Lund et al.(2003)]{2003A&A...411L.231L} Lund, N., Budtz-J{\o}rgensen, C., Westergaard, N.~J., et al.\ 2003, \aap, 411, L231 

\bibitem[Maan et al.(2016)]{2016ATel.8729....1M} Maan, Y., Surnis, M., Krishnakumar, M.~A., Joshi, B.~C., \& Manoharan, P.~K.\ 2016, The Astronomer's Telegram, 8729,  

\bibitem[Maeda et al.(2002)]{2002ApJ...570..671M} Maeda, Y., Baganoff, F.~K., Feigelson, E.~D., et al.\ 2002, \apj, 570, 671 

\bibitem[Masetti et al.(2016)]{2016ATel.8737....1M} Masetti, N., Saito, R.~K., Rojas, A.~F., \& Minniti, D.\ 2016, The Astronomer's Telegram, 8737,  

\bibitem[Miller et al.(2015)]{2015ApJ...814...87M} Miller, J.~M., Fabian, A.~C., Kaastra, J., et al.\ 2015, \apj, 814, 87 

\bibitem[Miralda-Escud{\'e} \& Gould(2000)]{2000ApJ...545..847M} Miralda-Escud{\'e}, J., \& Gould, A.\ 2000, \apj, 545, 847 

\bibitem[Mori et al.(2013)]{2013ApJ...770L..23M} Mori, K., Gotthelf, E.~V., Zhang, S., et al.\ 2013, \apjl, 770, L23 

\bibitem[Morris(1993)]{1993ApJ...408..496M} Morris, M.\ 1993, \apj, 408, 496 

\bibitem[Muno et al.(2005)]{2005ApJ...622L.113M} Muno, M.~P., Pfahl, E., Baganoff, F.~K., et al.\ 2005, \apjl, 622, L113 

\bibitem[Mu{\~n}oz-Darias et al.(2014)]{2014MNRAS.443.3270M} Mu{\~n}oz-Darias, T., Fender, R.~P., Motta, S.~E., \& Belloni, T.~M.\ 2014, \mnras, 443, 3270 

\bibitem[Mu{\~n}oz-Darias et al.(2011)]{2011MNRAS.410..679M} Mu{\~n}oz-Darias, T., Motta, S., \& Belloni, T.~M.\ 2011, \mnras, 410, 679 

\bibitem[Nandra et al.(1997)]{1997ApJ...476...70N} Nandra, K., George, I.~M., Mushotzky, R.~F., Turner, T.~J., \& Yaqoob, T.\ 1997, \apj, 476, 70 

\bibitem[Narayan et al.(1997)]{1997ApJ...478L..79N} Narayan, R., Garcia, M.~R., \& McClintock, J.~E.\ 1997, \apjl, 478, L79 

\bibitem[Neilsen \& Lee(2009)]{2009Natur.458..481N} Neilsen, J., \& Lee, J.~C.\ 2009, \nat, 458, 481 

\bibitem[Nowak et al.(2012)]{2012ApJ...759...95N} Nowak, M.~A., Neilsen, J., Markoff, S.~B., et al.\ 2012, \apj, 759, 95 

\bibitem[Pedicelli et al.(2009)]{2009A&A...504...81P} Pedicelli, S., Bono, G., Lemasle, B., et al.\ 2009, \aap, 504, 81 

\bibitem[Pfahl \& Loeb(2004)]{2004ApJ...615..253P} Pfahl, E., \& Loeb, A.\ 2004, \apj, 615, 253 

\bibitem[Plotkin et al.(2012)]{2012MNRAS.419..267P} Plotkin, R.~M., Markoff, S., Kelly, B.~C., K{\"o}rding, E., \& Anderson, S.~F.\ 2012, \mnras, 419, 267 

\bibitem[Pinto et al.(2010)]{2010A&A...521A..79P} Pinto, C., Kaastra, J.~S., Costantini, E., \& Verbunt, F.\ 2010, \aap, 521, A79 

\bibitem[Pons \& Rea(2012)]{2012ApJ...750L...6P} Pons, J.~A., \& Rea, N.\ 2012, \apjl, 750, L6 

\bibitem[Ponti et al.(2015)]{2015arXiv151008902P} Ponti, G., Bianchi, S., Munoz-Darias, T., et al.\ 2015, arXiv:1510.08902 

\bibitem[Ponti et al.(2015)]{2015MNRAS.446.1536P} Ponti, G., Bianchi, S., Mu{\~n}oz-Darias, T., et al.\ 2015, \mnras, 446, 1536 

\bibitem[Ponti et al.(2015)]{2015MNRAS.454.1525P} Ponti, G., De Marco, B., Morris, M.~R., et al.\ 2015, \mnras, 454, 1525 

\bibitem[Ponti et al.(2015)]{2015MNRAS.453..172P} Ponti, G., Morris, M.~R., Terrier, R., et al.\ 2015, \mnras, 453, 172 

\bibitem[Ponti et al.(2014)]{2014MNRAS.444.1829P} Ponti, G., Mu{\~n}oz-Darias, T., \& Fender, R.~P.\ 2014, \mnras, 444, 1829 

\bibitem[Ponti et al.(2014)]{2014IAUS..303..333P} Ponti, G., Morris, M.~R., Clavel, M., et al.\ 2014b, The Galactic Center: Feeding and Feedback in a Normal Galactic Nucleus, 303, 333 

\bibitem[Ponti et al.(2013)]{2013ASSP...34..331P} Ponti, G., Morris, M.~R., Terrier, R., \& Goldwurm, A.\ 2013, Cosmic Rays in Star-Forming Environments, 34, 331 

\bibitem[Ponti et al.(2012)]{2012MNRAS.422L..11P} Ponti, G., Fender, R.~P., Begelman, M.~C., et al.\ 2012, \mnras, 422, 11 

\bibitem[Ponti et al.(2010)]{2010ApJ...714..732P} Ponti, G., Terrier, R., Goldwurm, A., Belanger, G., \& Trap, G.\ 2010, \apj, 714, 732 

\bibitem[Ponti et al.(2004)]{2004A&A...417..451P} Ponti, G., Cappi, M., Dadina, M., \& Malaguti, G.\ 2004, \aap, 417, 451 

\bibitem[Predehl \& Schmitt(1995)]{1995A&A...293..889P} Predehl, P., \& Schmitt, J.~H.~M.~M.\ 1995, \aap, 293,  

\bibitem[Rea et al.(2011)]{2011ApJ...729L..21R} Rea, N., Jonker, P.~G., Nelemans, G., et al.\ 2011, \apjl, 729, L21 

\bibitem[Rea \& Esposito(2011)]{2011ASSP...21..247R} Rea, N., \& Esposito, P.\ 2011, Astrophysics and Space Science Proceedings, 21, 247 

\bibitem[Rea et al.(2013)]{2013ApJ...775L..34R} Rea, N., Esposito, P., Pons, J.~A., et al.\ 2013, \apjl, 775, L34 

\bibitem[Remillard \& McClintock(2006)]{2006ARA&A..44...49R} Remillard, R.~A., \& McClintock, J.~E.\ 2006, \araa, 44, 49 

\bibitem[Revnivtsev et al.(2004)]{2004A&A...425L..49R} Revnivtsev, M.~G., Churazov, E.~M., Sazonov, S.~Y., et al.\ 2004, \aap, 425, L49 

\bibitem[Reynolds et al.(2016)]{2016ATel.8649....1R} Reynolds, M., Kennea, J., Degenaar, N., Wijnands, R., \& Miller, J.\ 2016, The Astronomer's Telegram, 8649,  

\bibitem[Smith et al.(2016)]{2016ApJ...818..143S} Smith, R.~K., Valencic, L.~A., \& Corrales, L.\ 2016, \apj, 818, 143 

\bibitem[Str{\"u}der et al.(2001)]{2001A&A...365L..18S} Str{\"u}der, L., Briel, U., Dennerl, K., et al.\ 2001, \aap, 365, L18 

\bibitem[Terrier et al.(2010)]{2010ApJ...719..143T} Terrier, R., Ponti, G., B{\'e}langer, G., et al.\ 2010, \apj, 719, 143 

\bibitem[Turner et al.(2001)]{2001A&A...365L..27T} Turner, M.~J.~L., Abbey, A., Arnaud, M., et al.\ 2001, \aap, 365, L27 

\bibitem[Turolla et al.(2015)]{2015RPPh...78k6901T} Turolla, R., Zane, S., \& Watts, A.~L.\ 2015, Reports on Progress in Physics, 78, 116901 

\bibitem[Ubertini et al.(2003)]{2003A&A...411L.131U} Ubertini, P., Lebrun, F., Di Cocco, G., et al.\ 2003, \aap, 411, L131 

\bibitem[van der Klis(1988)]{1988tns..conf...27V} van der Klis, M.\ 1988, Timing Neutron Stars, eds.~H.~Ogelman and E.P.J.~van den Heuvel.~ NATO ASI Series C, Vol.~262, p.~27-70.~Dordrecht: Kluwer, 1988., 262, 27 

\bibitem[Vaughan et al.(1994)]{1994ApJ...435..362V} Vaughan, B.~A., van der Klis, M., Wood, K.~S., et al.\ 1994, \apj, 435, 362 

\bibitem[Vaughan et al.(2003)]{2003MNRAS.345.1271V} Vaughan, S., Edelson, R., Warwick, R.~S., \& Uttley, P.\ 2003, \mnras, 345, 1271 

\bibitem[Verner et al.(1996)]{1996ApJ...465..487V} Verner, D.~A., Ferland, G.~J., Korista, K.~T., \& Yakovlev, D.~G.\ 1996, \apj, 465, 487 

\bibitem[Wallace \& Hinkle(1997)]{1997ApJS..111..445W} Wallace, L., \& Hinkle, K.\ 1997, \apjs, 111, 445 

\bibitem[Walter et al.(2015)]{2015A&ARv..23....2W} Walter, R., Lutovinov, A.~A., Bozzo, E., \& Tsygankov, S.~S.\ 2015, A\&ARv, 23, 2 

\bibitem[Wharton et al.(2012)]{2012ApJ...753..108W} Wharton, R.~S., Chatterjee, S., Cordes, J.~M., Deneva, J.~S., \& Lazio, T.~J.~W.\ 2012, \apj, 753, 108 

\bibitem[White \& Mason(1985)]{1985SSRv...40..167W} White, N.~E., \& Mason, K.~O.\ 1985, \ssr, 40, 167 

\bibitem[Wijnands et al.(2006)]{2006A&A...449.1117W} Wijnands, R., in't Zand, J.~J.~M., Rupen, M., et al.\ 2006, \aap, 449, 1117 

\bibitem[Wilms et al.(2000)]{2000ApJ...542..914W} Wilms, J., Allen, A., \& McCray, R.\ 2000, \apj, 542, 914 

\bibitem[Wilms et al.(2010)]{2010HEAD...11.1206W} Wilms, J., Lee, J.~C., Nowak, M.~A., et al.\ 2010, Bulletin of the American Astronomical Society, 42, 12.06 

\bibitem[Zhang et al.(2015)]{2015ApJ...815..132Z} Zhang, S., Hailey, C.~J., Mori, K., et al.\ 2015, \apj, 815, 132 

\bibitem[Zdziarski et al.(1996)]{1996MNRAS.283..193Z} Zdziarski, A.~A., Johnson, W.~N., \& Magdziarz, P.\ 1996, \mnras, 283, 193 

\bibitem[{\.Z}ycki et al.(1999)]{1999MNRAS.309..561Z} {\.Z}ycki, P.~T., Done, C., \& Smith, D.~A.\ 1999, \mnras, 309, 561 

\end{thebibliography}
\end{document}